\documentclass[useAMS,usenatbib]{mn2e}
\usepackage{graphicx,natbib,times,mathptm}

\def\Halpha{{\rm H}\alpha}

\bibpunct{(}{)}{;}{a}{}{,}

\begin{document}

\bibliographystyle{mn2e}

\title[Extinction Mapping With IPHAS]{High Spatial Resolution Galactic
3D Extinction Mapping with IPHAS}

\author[Sale et al.]{Stuart E. Sale$^1$\thanks{E-mail: s.sale06@imperial.ac.uk}, J. E. Drew$^{1,2}$, Y. C. Unruh$^1$, M.J. Irwin$^3$, C. Knigge$^4$, S. Phillipps$^{5}$,
\newauthor A. A. Zijlstra$^{6}$, B. T. G\"ansicke$^7$, R. Greimel$^{8}$, P. J. Groot$^{9}$, A. Mampaso$^{10}$,
\newauthor  R. A. H. Morris$^{5}$,  R. Napiwotzki$^2$, D. Steeghs$^{7,11}$ N. A. Walton$^{3}$\\
$^1$ Astrophysics Group, Imperial College London, Blackett Laboratory, Prince Consort Road, London, SW7~2AZ, U.K.\\
$^2$ Centre for Astrophysics Research, STRI, University of Hertfordshire, College Lane Campus, Hatfield, AL10~9AB, U.K.\\
$^3$ Institute of Astronomy, Madingley Road, Cambridge CB3~0HA, U.K.\\
$^4$ School of Physics \& Astronomy, University of Southampton, Southampton, SO17 1BJ, U.K. \\
$^5$ Astrophysics Group, Department of Physics, Bristol University, Tyndall Avenue, Bristol, BS8 1TL, U.K.\\
$^6$ Jodrell Bank Center for Astrophysics, Alan Turing Building, The University of Manchester, Oxford Street, Manchester, M13~9PL, U.K. \\
$^7$ Department of Physics, University of Warwick, Coventry, CV4 7AL, U.K. \\
$^8$ Institut f\"ur Physik, Karl-Franzens Universit\"at Graz, Universit\"atsplatz~5, 8010 Graz, Austria\\
$^9$ Department of Astrophysics/IMAPP, Radboud University Nijmegen, P.O. Box 9010, 6500 GL, Nijmegen, The Netherlands \\
$^{10}$ Instituto de Astrof\'{\i}sica de Canarias, Via L\'actea s/n, E38200 La Laguna, Santa Cruz de Tenerife, Spain \\
$^{11}$ Harvard-Smithsonian Center for Astrophysics, 60 Garden Street, Cambridge, MA 
02138, USA\\
}

\date{Received .........., Accepted...........}

\maketitle

\begin{abstract}
We present an algorithm ({\scshape mead}, for `Mapping Extinction Against Distance') which will determine intrinsic ($r' - i'$) colour, extinction, and distance for early-A to K4 stars extracted from the IPHAS $r'/i'/\Halpha$ photometric database.  These data can be binned up to map extinction in three dimensions across the northern Galactic Plane. The large size of the IPHAS database ($\sim 200$ million unique objects), the accuracy of the digital photometry it contains and its faint limiting magnitude ($r' \sim 20$) allow extinction to be mapped with fine angular ($ \sim 10 $ arcmin) and distance ($\sim 0.1$ ~kpc) resolution to distances of up to $10$ kpc, outside the Solar Circle. High reddening within the Solar Circle on occasion brings this range down to $\sim 2$~kpc. The resolution achieved, both in angle and depth, greatly exceeds that of previous empirical 3D extinction maps, enabling the structure of the Galactic Plane to be studied in increased detail. {\scshape mead} accounts for the effect of the survey magnitude limits, photometric errors, unresolved ISM substructure, and binarity. The impact of metallicity variations, within the range typical of the Galactic disc is small. The accuracy and reliability of {\scshape mead} are tested through the use of simulated photometry created with Monte-Carlo sampling techniques. The success of this algorithm is demonstrated on a selection of fields and the results are compared to the literature. 
\end{abstract}

\section{Introduction}

The INT/WFC Photometric $\Halpha$ Survey of the northern Galactic Plane \citep[IPHAS; ][]{Drew.2005short} is the first comprehensive digital survey of the disc of the Galaxy ($|b|\leq5^{\circ}$), north of the celestial equator. Imaging is performed in the $r'$, $i'$ and $\Halpha$ bands down to $r'\sim20$ ($10\sigma$). The high quality photometry from this survey has been shown to enable early-A type stars to be selected very easily, due to their strong $\Halpha$ absorption \citep{Drew.2008}. In this same study, the $(r' - i')$ colours provided by IPHAS were used to determine the extinctions of the A stars selected, and then distances were deduced using photometric parallax.  This capability is a specific example of a more general property of the IPHAS: since $(r' - \Halpha)$ correlates strongly with intrinsic ($r' - i'$) colour, the combination of $r'$, $i'$ and $\Halpha$ can be used to disentangle intrinsic colour and extinction unambiguously.  The aim of this paper is to present a method which extends the approach of \cite{Drew.2008} to incorporate a broader range of spectral types and, therefore, many more objects. Then extinction can be successfully mapped at fine angular resolutions out to large distances.

The IPHAS catalogue contains high quality photometry for $\sim200$ million objects, equating to over $10^5$ objects per square degree. Our analysis of a wide variety of IPHAS fields has shown that the data quality allows for accurate determinations of distances and extinctions to the studied objects, with median errors of typically $\sim 100$pc on distance and approximately $0.1$ magnitudes of error on the determined extinction. The limiting magnitude of the IPHAS catalogue ($r'\sim20, 10\sigma$) is sufficient for extinctions to be measured to distances up to $\sim 10$~kpc from the Sun.

\cite{Juric.2008short} discuss and demonstrate the potential for accurate wide area surveys in analysing the structure of the Milky Way. They utilise the Sloan Digital Sky Survey (SDSS) in order to determine the stellar number density distribution mostly at Galactic latitudes $|b| > 25^{\rm o}$. As a result of this they are able to easily study the Galactic halo and local scale heights of the Milky Way's thick and thin discs, and find localised over-densities. In contrast the IPHAS survey area covers low Galactic latitudes, enabling the Galactic thin disc to be studied in detail, well beyond the solar neighbourhood. The cost associated with this is that extinction, which is largely negligible in the SDSS survey area, becomes a significant concern to the extent that understanding its three-dimensional (3D) distribution is a necessary hurdle that must be overcome on the path to studying the structure of the Galactic disc.

Applications of extinction mapping go beyond studies of Galactic structure: a distance-extinction relationship, combined with an estimate of the extinction to an object has often been exploited to give an estimate of the distance to individual objects. Such estimates are of particular use for objects such as planetary nebulae, for which reliable estimates of distance are difficult to obtain (Viironen et al. in preparation).

There is a long history associated with trying to map the distribution of extinction, going back as far as \cite{Trumpler.1930} and  \cite{Van_de_Kamp.1930} who both appreciated that extinction is concentrated near the Galactic plane. Subsequent efforts to map extinction can be broken down into two broad categories: 2D maps of the asymptotic extinction and 3D maps with varying degrees of model input.

Several studies have produced 2D maps of the maximum asymptotic extinction over the sky. \cite{Burstein.1982} used HI mapping and galaxy counts, estimating the errors in derived extinctions to be $\sim10\%$. \cite*{Schlegel.1998} provided updated maps based on COBE and FIRAS measurements of dust emission at $100\mu m$. The \cite{Schlegel.1998} maps have since been widely used, but they warn that their values for $|b|<5^{\circ}$ are subject to significant errors, an assertion that has been supported by \cite{Arce.1999} and \cite*{Cambresy.2005}. More recently \cite{Froebrich.2005} have used cumulative star counts from the 2MASS survey to construct a 2D extinction map of the Galactic plane ($|b|<20^{\circ}$).

3D extinction models have grown progressively more sophisticated, moving from the simple cosecant model \citep{Parenago.1945} to the modern, high resolution models of \cite{Mendez.1998, Chen.1999, Drimmel.2001} and \cite{Amores.2005}.  \cite{Marshall.2006} compared the Besan\c{c}on Galactic model \citep{Robin.2003}, without extinction, to 2MASS (Two Micron All-Sky Survey) observations. The differences between the observed and theoretical colours of predominantly K \& M giants were then attributed to extinction. This semi-empirical method is able to map extinction at fine angular resolutions ($0.25^{\circ} \times 0.25^{\circ}$).

Historically, attempts to empirically map extinction in three dimensions have been constrained by the availability and quality of photometric data. Early studies include \cite{Fernie.1962, Eggen.1963, Isserstedt.1964, Neckel.1966, Scheffler.1966, Scheffler.1967}, all of which reached no fainter than $V\sim10$ and therefore could only cover relatively limited ranges of up to a $1-2$~kpc. Also, they all contain fewer than 5000 objects, limiting their mappings to very coarse spatial resolutions. Furthermore, \cite{Fitzgerald.1968} found that these works produced `varying, and sometimes apparently conflicting results'. Subsequently, \cite{Fitzgerald.1968, Lucke.1978} and  \cite*{Neckel.1980} continued, with similar approaches, but they were still limited by the available photometric catalogues, with the most recent of the three \citep{Neckel.1980} using a catalogue of $\sim 11,000$ objects. As a result, they achieved angular resolutions on the order of a few degrees.

Here, we present a new approach to mapping the 3D extinction within the plane of our Galaxy, based on IPHAS photometry. The depth, quantity and quality of this data set allows us to construct a new extinction map with high angular ($10 \arcmin$) and distance (100 pc) resolution out to very large distances (up to 10 kpc). In the present paper, we focus on the algorithm used to construct this map, leaving the map itself to be presented in a separate publication. The algorithm which is used to calculate extinction distance-relationships from IPHAS data (hereafter referred to as {\scshape mead}, Mapper of Extinction Against Distance) is discussed in detail in Section~\ref{algorithm}. In order for it to be used with IPHAS data it is necessary to calibrate this translation of IPHAS photometry into derived quantities (Section~\ref{calibration}). Section~\ref{simulated} shows the use of simulated photometry to determine the values of some key parameters used in {\scshape mead} and demonstrate its accuracy in the face of a range of possible problems. Finally, Section~\ref{examples}, investigates the use of {\scshape mead} on real IPHAS data, showing the power of its approach.

\section{Decoding intrinsic colour, luminosity class and extinction from IPHAS photometry}\label{calibration}

\begin{figure}
\centering
\includegraphics[width=80mm, height=60mm]{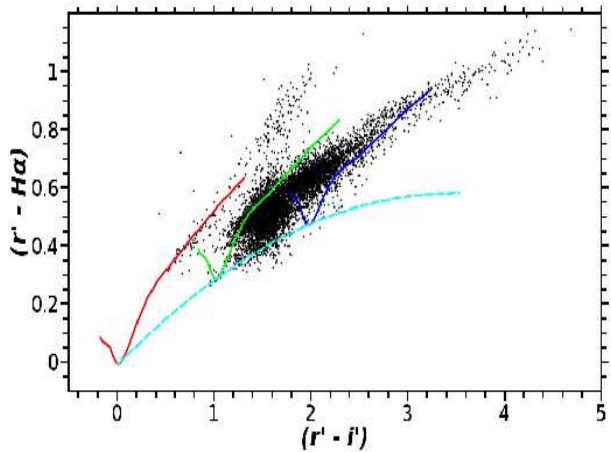}
\caption{Black points show data from IPHAS field 4199, where $13 \leq r' < 20$. Main sequences where extinctions equivalent to $A_{V}=0$ (red), $4$ (green), $8$ (blue) for an A0V star have been applied are shown with solid lines and the dashed cyan line shows an A3V reddening line. There are two main populations visible in Fig.~\ref{example_data}, the first lies at bluer ($r' - i'$) aligned with and extending beyond the simulated unreddened main sequence: this population is mainly K-M dwarfs. The main body of stars falls between the $A_V=4$ and $A_V=8$ main sequences. The long trail of objects stretching from ($r' - i'$)$ \approx 2$ are M giants. Finally, there are five objects that appear to lie above the unreddened main sequence, these objects are candidate $\Halpha$ emitters.\label{example_data}}
\end{figure}

In order to exploit IPHAS, it is necessary to understand the behaviour of its colour-colour and colour-magnitude planes -- specifically the intrinsic colours of normal stars and their response to extinction. To this end \cite{Drew.2005short} produced simulations of the luminosity class sequences on the IPHAS colour-colour plane using the spectral library of \cite{Pickles.1998}. In Fig.~\ref{example_data} we provide an example of $(r'-\Halpha, r' - i')$ data for a survey field with differently reddened simulated main sequences spanning the O9 to M3 spectral type range superposed:  this shows how reddening shifts the main sequence across to the right, creating a lower edge to the main stellar locus that is well-described as the early-A reddening line. Reddening also moves O and B stars to the same colours as A and F stars, though they are still separated in apparent magnitude. For the purpose of emulating the colour behaviour of normal stars, the \cite{Pickles.1998} library contains limited numbers of objects that cannot support a full exploration of the available parameter space.  In particular, in order to examine the impact of changing metallicity and fine variations in luminosity class, we have recomputed expected IPHAS colours from grids of model atmospheres.

New luminosity class sequences have been obtained for the IPHAS colour-colour and colour-magnitude planes using the \cite{Munari.2005} library of ATLAS9 synthetic spectra. \cite{Kurucz.2005} discusses some particular limitations of ATLAS9 models, some of which are relevant to this study: ATLAS9 spectra include only an incomplete sample of diatomic molecules and, with the exception of H$_{2}$O, no triatomic or larger molecules. This inherently leads to deviations from reality, which become significant for late K and all M type stars \citep{Bertone.2004}. In particular the absence of VO lines means that Kurucz models are not available for $T_{\rm eff}<3500K$ (approximately M4 or later) as they would be particularly inaccurate \citep{Kurucz.2005}.  Similar problems affect other grids of atmospheres, with the consequence that, for our purposes, there are no reliable alternative model grids at these low temperatures.

\subsection{Method and underpinning stellar data}

\begin{table}
\caption{The intrinsic ($r' - i'$) and ($r'-\Halpha$) colours, and r' magnitude compared to approximate spectral type and $T_{\rm eff}$ for solar metallicity dwarfs. The data are derived from the \protect\cite{Straizys.1981} $T_{\rm eff}$, $\log g$ calibration and the absolute magnitude calibration of \protect\cite{Houk.1997}. \label{type_colours}}
\begin{tabular}{ c c c c c}
 Intrinsic  & Intrinsic & $r'$ Absolute & Spectral & $T_{\rm eff}$ \\
 ($r' - i'$) & ($r'-\Halpha$) &Magnitude & Type & \\
\hline
-0.047	&	0.029	&	0.12	&	B8	&	11510	\\
-0.028	&	0.013	&	0.51	&	B9	&	10400	\\
-0.006	&	-0.002	&	0.80	&	A0	&	9600	\\
0.001	&	-0.005	&	1.09	&	A1	&	9400	\\
0.013	&	-0.008	&	1.28	&	A2	&	9150	\\
0.025	&	-0.008	&	1.47	&	A3	&	8900	\\
0.060	&	0.006	&	1.84	&	A5	&	8400	\\
0.097	&	0.032	&	2.21	&	A7	&	8000	\\
0.171	&	0.098	&	2.64	&	F0	&	7300	\\
0.203	&	0.128	&	2.92	&	F2	&	7000	\\
0.258	&	0.174	&	3.37	&	F5	&	6500	\\
0.296	&	0.202	&	3.83	&	F8	&	6150	\\
0.318	&	0.218	&	4.11	&	G0	&	5950	\\
0.336	&	0.229	&	4.39	&	G2	&	5800	\\
0.372	&	0.252	&	4.75	&	G5	&	5500	\\
0.406	&	0.269	&	5.21	&	G8	&	5250	\\
0.436	&	0.283	&	5.58	&	K0	&	5050	\\
0.454	&	0.291	&	5.76	&	K1	&	4950	\\
0.472	&	0.298	&	6.04	&	K2	&	4850	\\
0.504	&	0.312	&	6.20	&	K3	&	4700	\\
0.527	&	0.322	&	6.48	&	K4	&	4600	\\
\end{tabular}
\end{table}

\begin{figure}
\centering
\includegraphics[width=80mm, height=60mm]{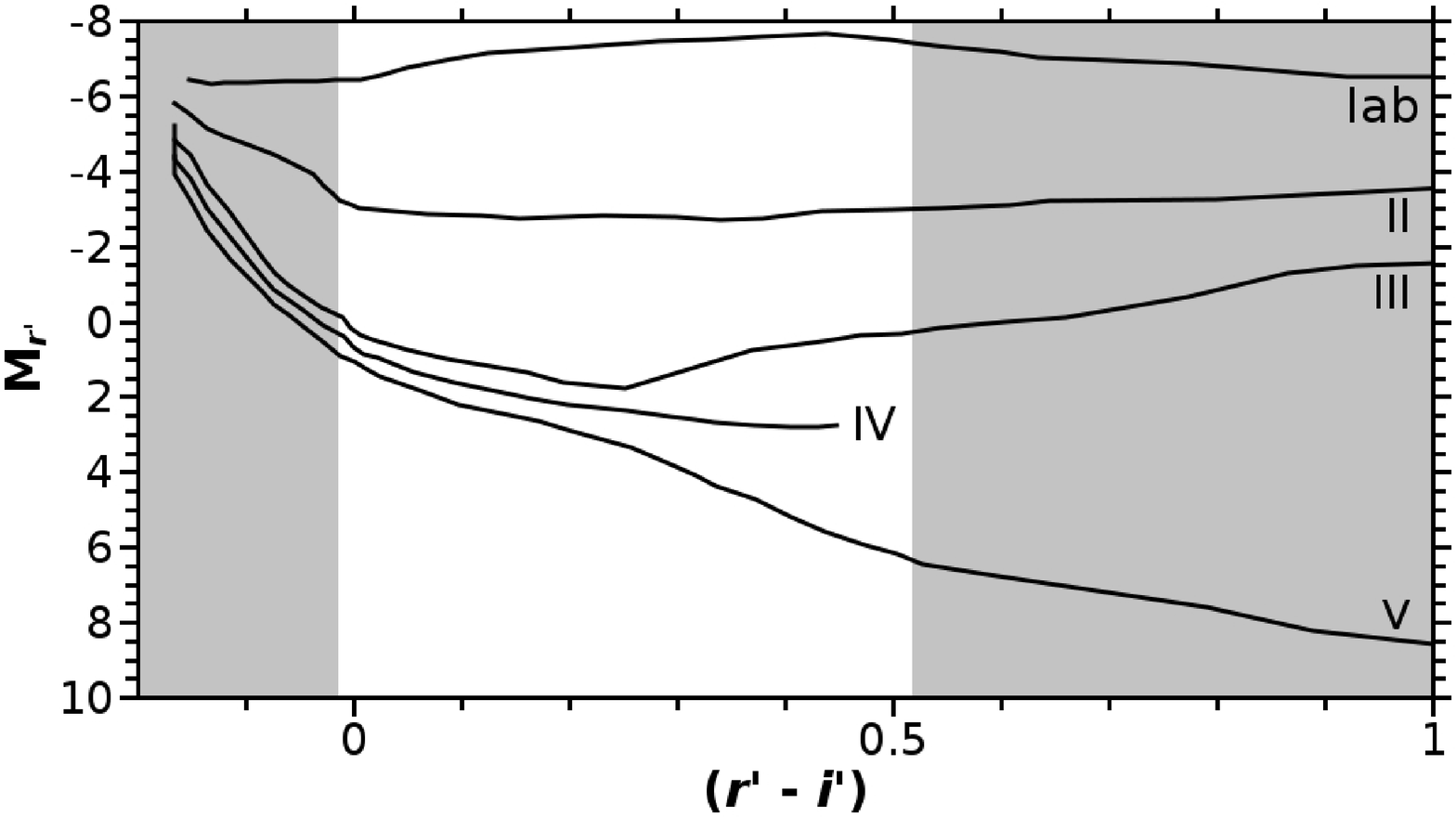}
\caption{A Hertzsprung-Russell diagram of the simulated luminosity classes used in this study. The unshaded region of the diagram indicates the intrinsic colour range of interest, which is the range of intrinsic colour that {\scshape mead} works on. \label{HRD}}
\end{figure}

So as to use the synthetic spectral libraries to create luminosity class sequences, the classes of the MK system \citep*{Morgan.1943} must be linked to $T_{\rm eff}$, $\log g$ and absolute magnitudes ($M_{V}$). This study adopts the $T_{\rm eff}$ and $\log g$ calibrations of \cite{Straizys.1981}, as well as their absolute magnitude calibrations for all classes except dwarfs, for which we use the data of \cite{Houk.1997} based on HIPPARCOS parallaxes. In Table~\ref{type_colours} we give the solar-metallicity mapping of intrinsic $(r' - i')$ colour onto main sequence absolute magnitude and spectral type that we have adopted, while Fig.~\ref{HRD} shows the relevant Hertzsprung-Russell diagram of the five luminosity classes we use.   

All synthetic colours were determined largely as by \cite{Drew.2005short}, with the modest difference that the atmospheric transmission curve at the INT is included in the calculation, so that the total transmission profile is the product of the atmospheric transmission, CCD response and filter transmission curves.

As IPHAS magnitudes are based on the Vega system, an appropriate spectrum of Vega was required to provide the zero-point flux. So as to match spectral resolution (which is necessary to produce accurate $\Halpha$ magnitudes) the ATLAS9 Vega model spectrum available via \textit{kurucz.harvard.edu} was used. This assumes $T_{\rm eff}=9550K$ and $\log g= 3.80$. It should be noted that there has been much debate over which ATLAS9 spectra best fits observations of Vega. For example, \cite{Cohen.1992} and \cite{Bohlin.2007} argue that the 9400K model (as opposed to the 9550K model used in this work) is better suited. However, the differences in derived colours are small, $\delta$($r' - \Halpha$)$\approx$ 0.003, so the choice from these two models is of little consequence.

For comparison to the synthetic sequences derived in this study, sequences based on libraries of intrinsic (i.e unreddened or dereddened) empirical  SEDs are employed. At the time of writing there are over 100 electronic libraries available, summarised in the Asiago Database of Spectrophotometric Databases \citep[\textit{http://web.pd.astro.it/adsd}]{Sordo.2006}. However, of these only a small proportion cover the wavelength range required and many of those are unsuitable. Once these have been filtered out, the STELIB library \citep{LeBorgne.2003} and the library of \cite{Pickles.1998} remain. 

\subsection{Comparisons between colour sequences from synthetic and empirical libraries}

\begin{figure}
\centering
\includegraphics[width=80mm, height=60mm]{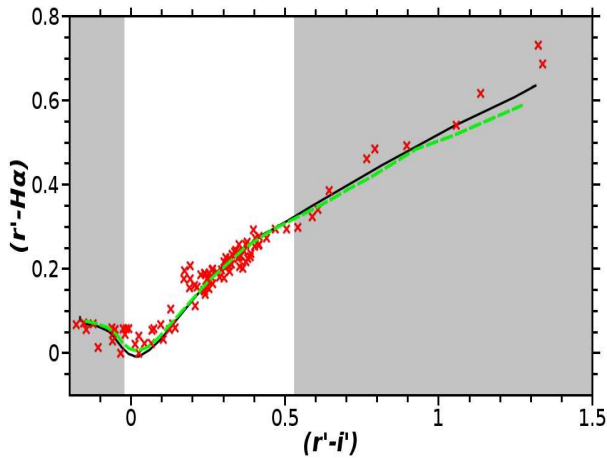}
\caption{The simulated main (solid black line) and giant (dashed green line) sequences compared to solar metallicity main sequence objects from the \protect\cite{Pickles.1998} and STELIB \protect\citep{LeBorgne.2003} libraries (red crosses). As with figure \protect\ref{HRD}, the unshaded region of the diagram indicates the intrinsic colour range of interest, used by {\scshape MEAD}.\label{empiricalcompare}}
\end{figure}

Fig.~\ref{empiricalcompare} compares the Main Sequence derived from the synthetic spectra of \cite{Munari.2005}, to colours derived from empirical spectra.  For most spectral types the results from empirical and synthetic spectra agree well. However, for M-type stars there is some disagreement. This could be a result of difficulties associated with modelling late type stars, as noted earlier. A comparison was also performed to sequences derived from the MARCS \citep{Gustafsson.2003} and PHOENIX \citep{Brott.2005} codes. The MARCS library only covers spectral types from early-A to early-K, but it is in excellent agreement with the results obtained from \cite{Munari.2005}. On the other hand, the sequences of \cite{Brott.2005} are significantly different for spectral types later than roughly K4. On further examination the differences are almost entirely limited to behaviour in the $\Halpha$ band. It is worth noting that \cite{Gustafsson.2008} demonstrate that a new version of the PHOENIX code, employing the Kurucz line data, performs considerably better than the version employed here. Therefore it is possible that the discrepancies we encounter will not be present in this newer version.

\cite{Bertone.2004} finds the ATLAS9 and PHOENIX models which best fit intrinsic spectra of known type and list the $T_{\rm eff}$ of the fit. By assuming that the $T_{\rm eff}$ calibration of \cite{Straizys.1981} represents the mean temperature for any given class, it is possible to calculate the standard deviation of the residuals of the \cite{Bertone.2004} estimates. This is then an estimate of the combination of the error on the \cite{Straizys.1981} $T_{\rm eff}$ calibration and the natural width of each class. However, allowing for this temperature variation makes no measurable difference to our model sequences.

Unfortunately \cite{Bertone.2004} do not give the values of $\log g$ for each fit, so it is not possible to repeat this procedure for surface gravity. However, \cite{Munari.2005} suggest that the natural width of each class in surface gravity is $\sim 0.25$ dex. Applying this spread results in the luminosity class sequences changing from lines to bands in colour-colour and colour-magnitude space. The majority of objects from the \cite{Pickles.1998} and STELIB libraries now fall within these bands. \cite{Bertone.2004} also demonstrate that the quality of the best fit to the stars in the empirical libraries it uses, decreases for later types. This behaviour is again attributable to the poorly understood molecular lines, which dominate the spectra of late-type stars.

\subsection{The impact of rotational broadening}
 
As might be expected, altering the rotational velocity assumed in the models has negligible effect on the colours. This is as varying rotational velocity will only alter the shape of a non-saturated line and will not affect the strength of absorption or emission. Because all the filters used are broad compared to emission or absorption lines, simply altering the line shapes does not change the observed colours. In saturated lines the equivalent width is affected by rotational broadening, but $\Halpha$ is the only atomic line which produces a significant impact on the photometric colours and it does not saturate, even in A3V stars where absorption is strongest. 

\begin{figure}
\includegraphics[width=80mm, height=60mm]{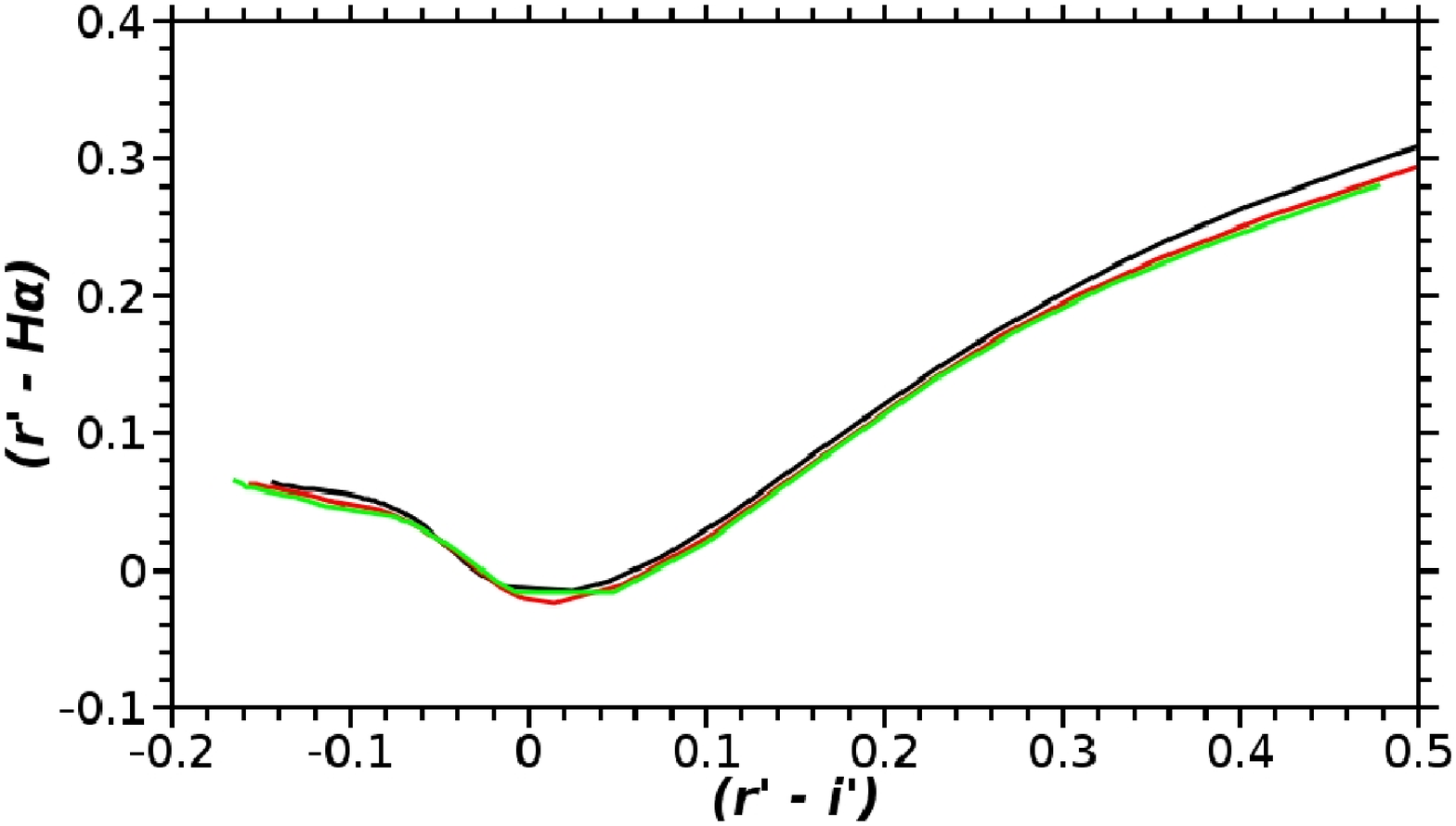}
\caption{ZAMS from the \protect\cite{Pietrinferni.2004} library on the colour-colour plane, where the colours have been determined from the appropriate \protect\cite{Munari.2005} models. In black is a ZAMS for [Fe/H]$=0$, in red a ZAMS for [Fe/H]$=-1.00$ and in green a ZAMS for [Fe/H]=$-1.82$. \label{metal_cc}}
\end{figure}

\begin{figure}
\centering
\includegraphics[width=80mm, height=60mm]{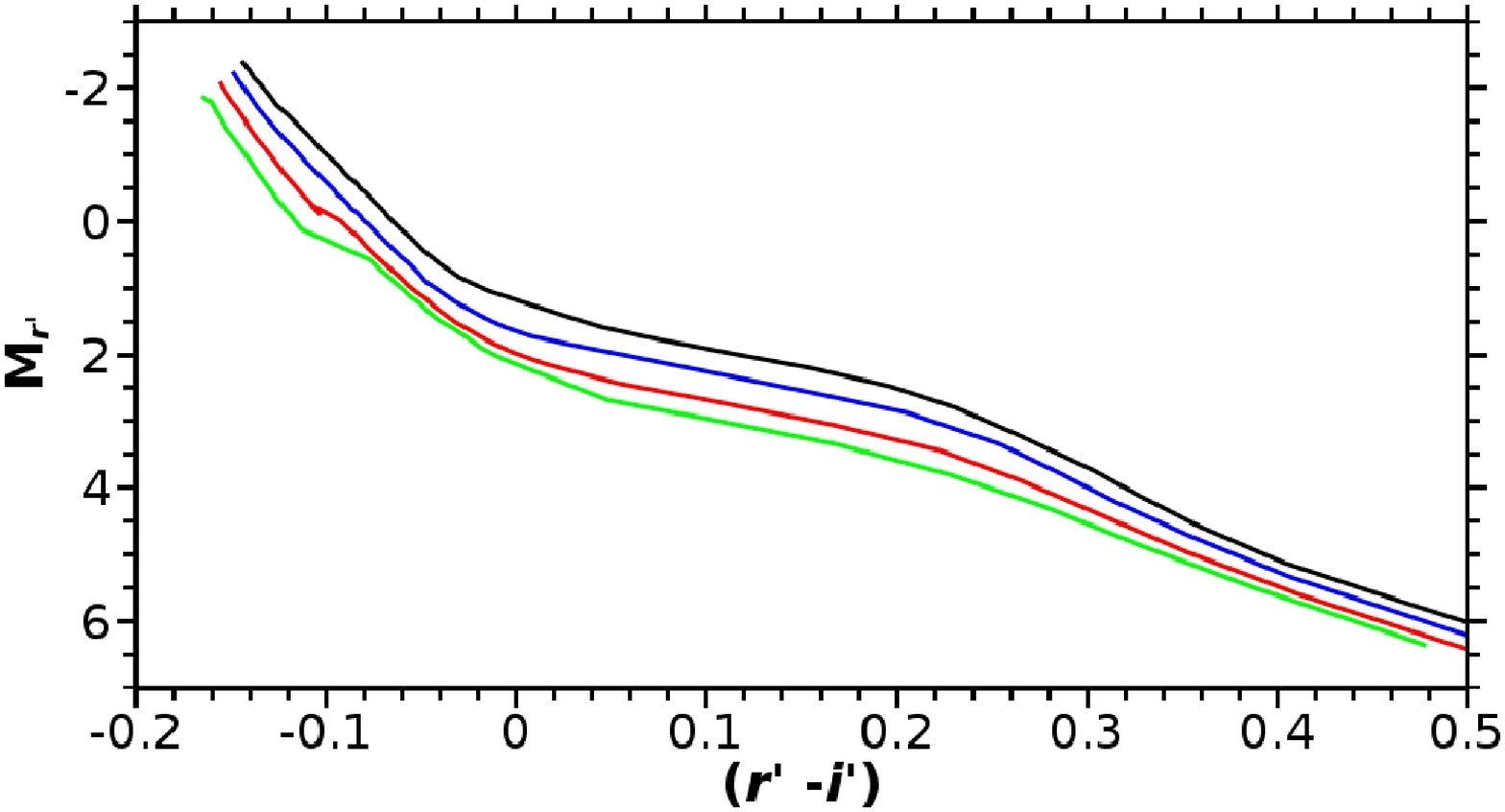}
\caption{ZAMS from the \protect\cite{Pietrinferni.2004} library on the colour-magnitude plane, where the colours have been determined from the appropriate \protect\cite{Munari.2005} models. The top, in black, is a ZAMS for [Fe/H]$=0$. The curve is lowered with decreasing metallicity, in blue is [Fe/H]$=-0.40$, red [Fe/H]$=-1.00$ and at the bottom, in green, is [Fe/H]=$-1.82$.\label{metal_cm}}
\end{figure}

\subsection{The impact of metallicity variation}\label{metal_calib}

Given that \cite{Straizys.1981} do not consider the effect of metallicity it is clearly not possible to use the sequences based on it to study the effect of metallicity. However, the library of \cite{Pietrinferni.2004} does consider it and so is used in this section. For \cite{Munari.2005} models altering metallicity has little effect on IPHAS colour-colour diagrams, a shown by Fig. \ref{metal_cc}.

Metallicity can, however, have a significant effect on the absolute magnitude of all objects. Fig.~\ref{metal_cm} demonstrates this for a zero-age main sequence (ZAMS). As it is not possible to determine the metallicity of the object from IPHAS observations, it is therefore also not possible to uniquely determine the absolute magnitude of an object. 

\cite{Friel.2002} measured metallicities in low Galactic latitude open clusters, finding no objects with [Fe/H]$<-0.94$, whilst \cite{Carraro.2007} determine the mean metallicity in the outer disc to be [Fe/H]$\simeq-0.35$. So, observing an object with a metallicity below [Fe/H]$=-0.5$ in IPHAS would be unusual and metallicities as low as [Fe/H]$=-1.82$ would be extremely rare. Therefore, the effect of metallicity variations on absolute magnitude will be relatively small in IPHAS observations. Section~\ref{metal_test} further examines the impact of undiagnosed metallicity variations.

\begin{figure}
\centering
\includegraphics[width=80mm, height=60mm]{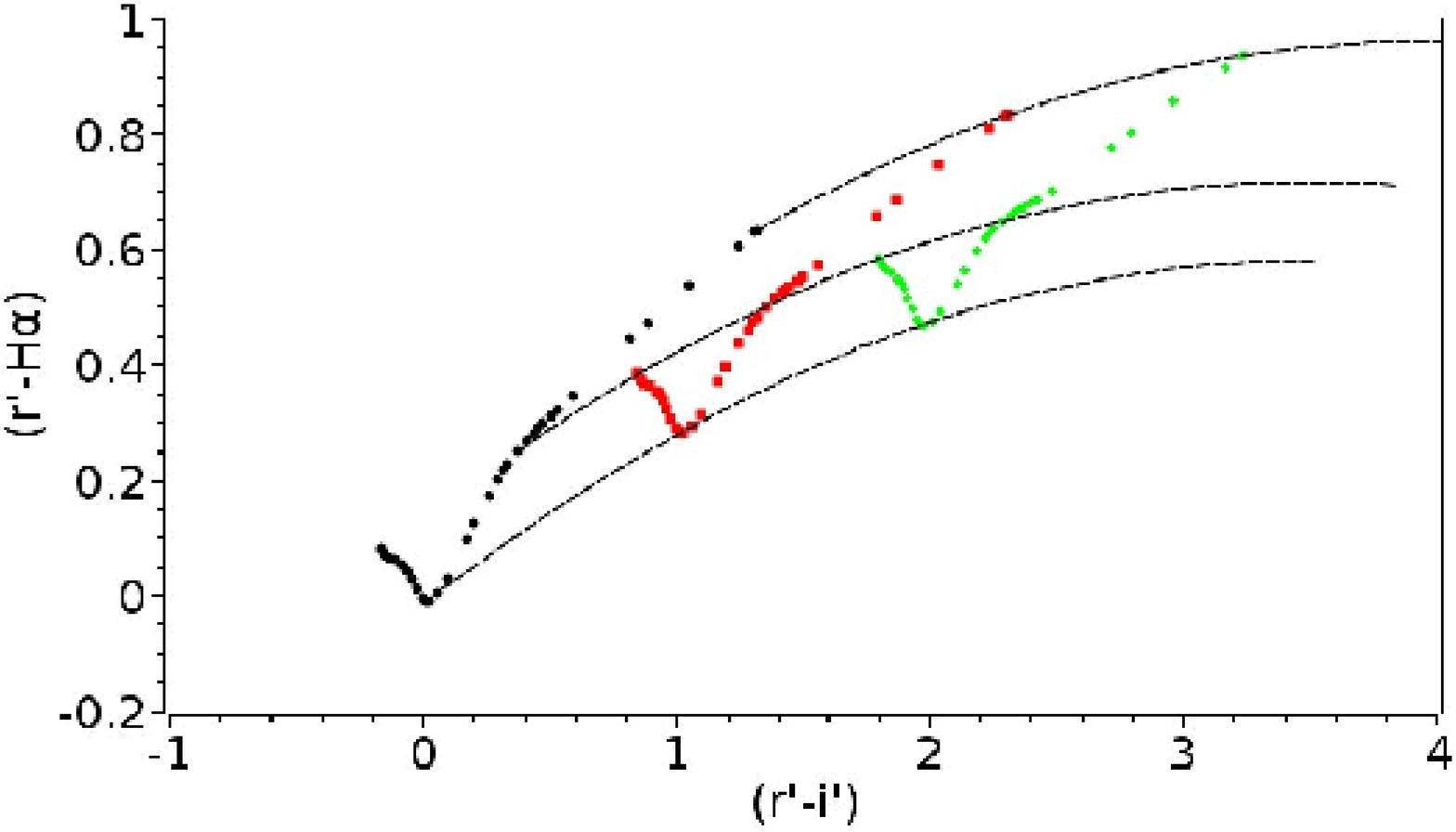}
\caption{Main sequences where extinctions equivalent to $A_{V}=0$ (black, left), $4$ (red, middle), $8$ (green, right) for an A0V star have been applied. The dashed black lines show the loci of A3V (bottom), G5V (middle) and M4V (top) stars under increasing extinction. \label{reddentracks}}
\end{figure}

\begin{figure}
\centering
\includegraphics[width=80mm, height=60mm]{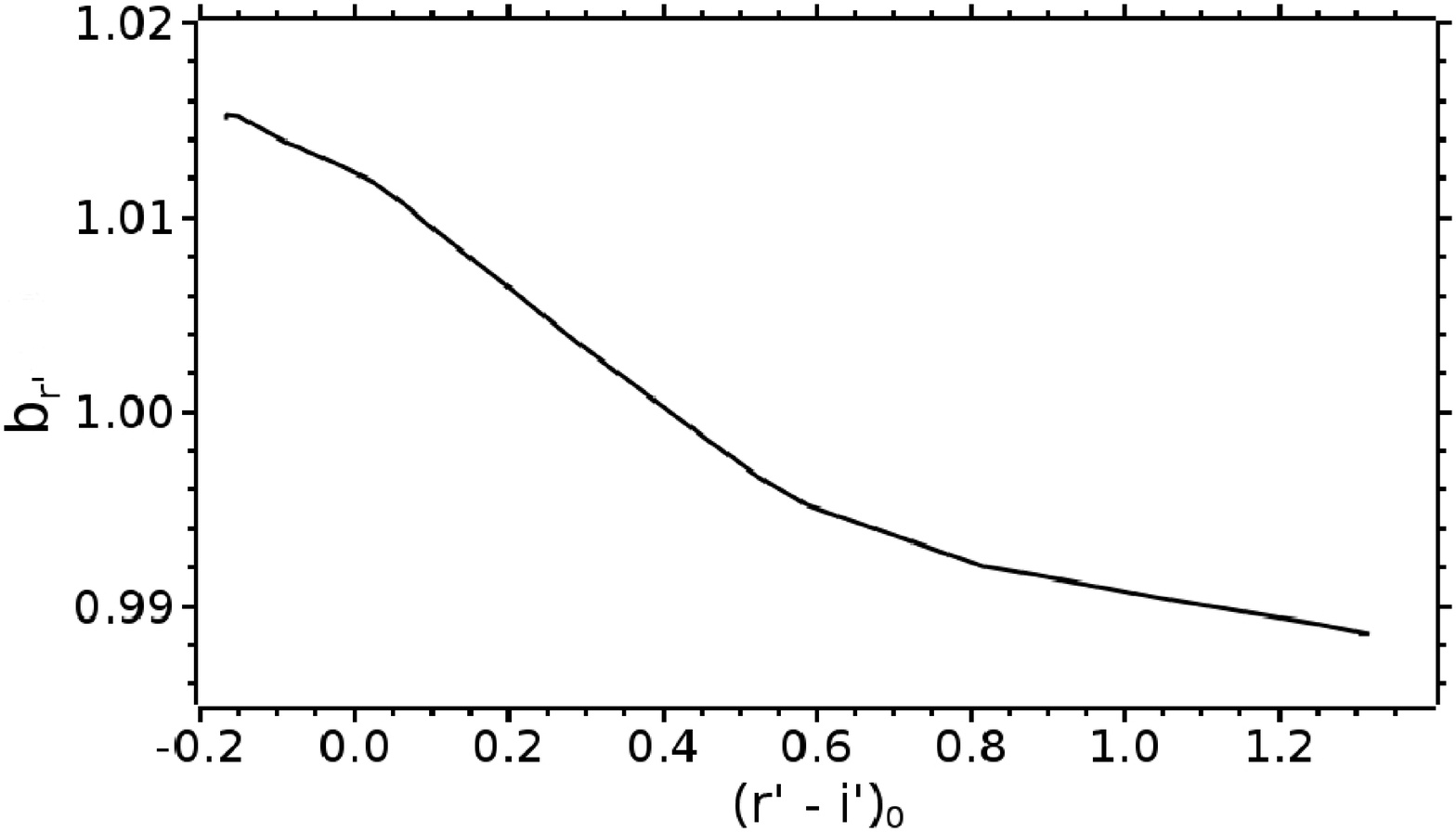}
\caption{The coefficient $b_{r'}$ against initial $(r'-i')$ for the \protect\cite{Straizys.1981} main sequence. Note that the coefficients $a_{r'}$ and $c_{r'}$, multiplying the squared and constant terms, are roughly $10^3$ times smaller than $b_{r'}$. \label{extcurves3}}
\end{figure}

\subsection{Extinction \& reddening}\label{ext_calib}

It is very important to note that, in the IPHAS colour-colour plane, the reddening vector is at a significant angle with respect to the majority of the main sequence (see Fig.~\ref{reddentracks}). This makes it possible to derive accurate estimates of intrinsic ($r' - i'$) colour and extinction simultaneously. For colours based on other commonly used filter sets such as $u'g'r'i'z'$ or $JHK$, the reddening vector is almost parallel to the main sequence, creating degeneracy in the available (intrinsic colour, extinction) solutions. This property of the IPHAS colour-colour plane is a result of the strong intrinsic ($r' - i'$) colour sensitivity of the ($r' - \Halpha$) colour, whilst its response to reddening is weak compared to that of broad-band colours. The former property is due to $(r' - \Halpha)$ being a proxy for the $\Halpha$ equivalent width, which itself is strongly dependent on intrinsic ($r' - i'$) colour. 

In many applications in astrophysics it is acceptable to treat the ratio $R_V=A_{V}/E(B-V)$ as constant across a wide range of objects, so that extinction can easily be calculated from reddening and vice-versa. However, as pointed out by \cite{McCall.2004} this is not in general true, as the extinction integrated across a given broadband filter is a function of the SED of the observed object and the amount and wavelength dependence of the extinction it has experienced. \cite{McCall.2004} suggests that monochromatic measures of extinction should be used, in preference to broadband measures, as they are independent of the observed objects' SEDs. Following this concept, {\scshape mead} measures a monochromatic extinction at $6250 \rm{\AA}$ ($A_{6250}$).  This is near the effective wavelength of the $r'$ filter in the INT system (i.e. taking into account the WFC and atmospheric response). If the intrinsic colour of the observed object is known it will then be possible to convert to and from more commonly-used broadband measures. In this study, we will refer to monochromatic extinctions in discussions of the simulated photometry, while extinctions derived from real data that are compared with the literature will be presented as broadband values, $A_{r'}$.  The extinction at $r'$ is preferred to the extinction in the $V$ band for the reasons discussed by \cite*{Cardelli.1989}. Section~\ref{law_test} investigates the ability of {\scshape mead} to change between monochromatic and broadband measures of extinction as $R$ is varied. By choosing to normalise the extinction law to a wavelength in the middle of the range studied the systematic error produced by uncertainty in the value of $R$ is minimised.

This work adopts the extinction laws of \cite{Fitzpatrick.2004} for the calculation of extinction values in different filters as a function of $A_{6250}$. Fortunately, this study avoids the problematic UV region where extinction laws are characterised by several variables, which are uncorrelated with $R$ \citep{Fitzpatrick.2007}. Here, the focus is on the IR/optical region where the wavelength dependence of extinction is well defined by $R$ only. Following \cite{Savage.1979} and \cite{Howarth.1983}, we assumed that $R=3.1$, so as to represent an average Galactic sight-line. 

The calibration of the broadband extinctions to $A_{6250}$ was performed using the \cite{Munari.2005} library to represent the \cite{Straizys.1981} sequences. Fitting was performed by linear regression of a quadratic polynomial to the data, over a range of $0\leq A_{6250}<10$, the functions returned, for a band X, are expressed as follows: $A_{X}=a_{X}A_{6250}^{2}+b_{X}A_{6250}+c_{X}$.

Fig.~\ref{reddentracks} demonstrates the effect of extinction on several different spectral types in the ($r'$ - $i', r'$ - $\Halpha$) plane. Despite being cosmetically similar, there are significant differences in the response to extinction of different spectral types. Fig.~\ref{extcurves3} demonstrates the variation of the coefficient for the leading linear term in the fits. For objects with types in the critical range A3 to K4 ($0.05\leq$($r'$ - $i'$)$<0.55$) there is a particularly simple linear relation between this fit coefficient and intrinsic ($r'$ -$i'$) colour, as can also be seen in Fig.~\ref{extcurves3}. 

The differing responses to rising extinction of different spectral types is illustrated by the extreme examples of A3V and M4V stars: in Fig.~\ref{reddentracks} these two types limit the plotted ($r'$ - $\Halpha$) range of the main sequence.  As more extinction is applied, the ($r'$ - $\Halpha$) range spanned shrinks because the A3V star is reddened more quickly in this colour.

\section{MEAD: an algorithm to determine extinction-distance relations}\label{algorithm}

\subsection{Concept}

The basic principle of this work is that we wish to determine the distance and intervening extinction to a large sample of objects in the IPHAS database. From the observations, we have three observed parameters for each object, namely the $r'$, $i'$ and $\Halpha$ magnitudes, with errors on each. In order to determine the distance to an object by photometric parallax, we must determine its absolute magnitude and extinction. The simplest way to estimate absolute magnitude is to determine the intrinsic colour and luminosity class of the object. Thus for each object we have three parameters we wish to determine (extinction, luminosity class \& intrinsic colour) and three observed parameters.

As described in Section~\ref{calibration}, it is possible to ascertain the spectral class and monochromatic extinction of an object from its position on the colour-colour plane. But, there are two main sources of degeneracy in this process. The first is caused by the fact that reddening tracks for O, B \& early A (A0-2) stars intersect those of later A and F stars. This follows from the fact that $\Halpha$ absorption peaks in early A stars and $(r' - \Halpha)$ is strongly correlated with $\Halpha$ equivalent width \citep{Drew.2005short}. As discussed in Section~\ref{contaminents_test}, it transpires that this is not normally a significant source of degeneracy.

\begin{figure}
\centering
\includegraphics[width=80mm, height=60mm]{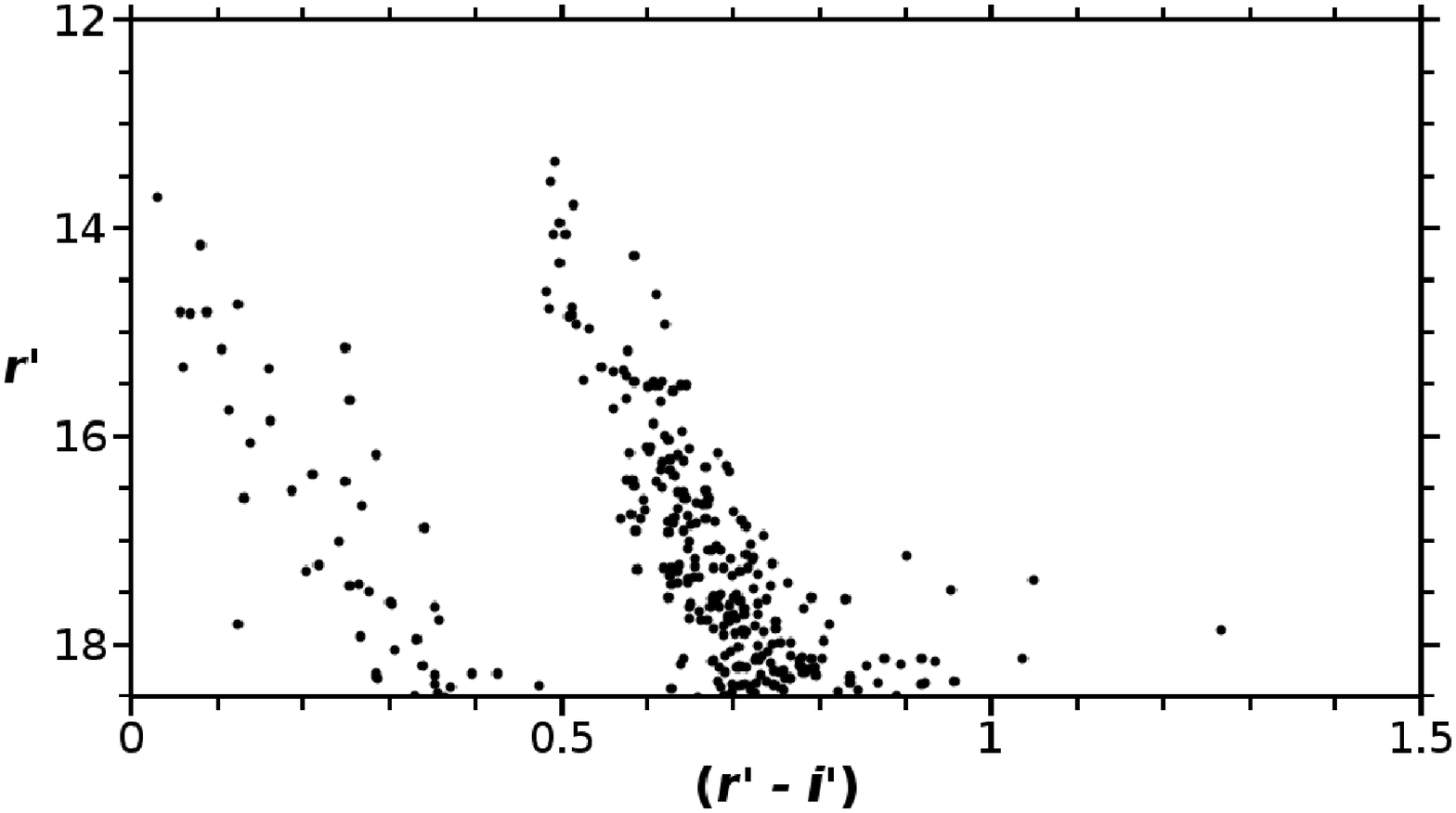}
\caption{A colour-magnitude diagram for F5-8 stars, selected by colour, from IPHAS observations of a $10 \arcmin \times 10 \arcmin$ box around $(l,b)=(101.55,-0.60)$ . The dwarfs can be seen as the sequence to the left, while the giants are seen to the right. Five brighter objects, probably either F bright-giants, or highly reddened late-O or early-B stars are visible to the extreme right, separated from the giant sequence. \label{lumclasser}}
\end{figure}

The second more important source of degeneracy is luminosity class, which has a significant impact on distance determination. Therefore, to estimate an object's distance, it is necessary for an estimate of the luminosity class to be determined first. This is, however, not a trivial task, as the different luminosity class sequences behave very similarly on the IPHAS colour-colour plane. We decide on the luminosity class of an object as follows. If two objects of identical intrinsic colours, but different absolute magnitude, are to have the same apparent magnitude the more luminous object must be further away, and is likely to be more heavily reddened.  This reddening contrast creates two sequences in a colour-magnitude diagram, as can be seen in Fig.~\ref{lumclasser}, where a selection of mainly F stars are clearly split into dwarf and giant sequences. It is by identifying these split sequences in the colour-magnitude diagram that luminosity class is assigned.

\begin{figure}
\centering
\includegraphics[width=80mm]{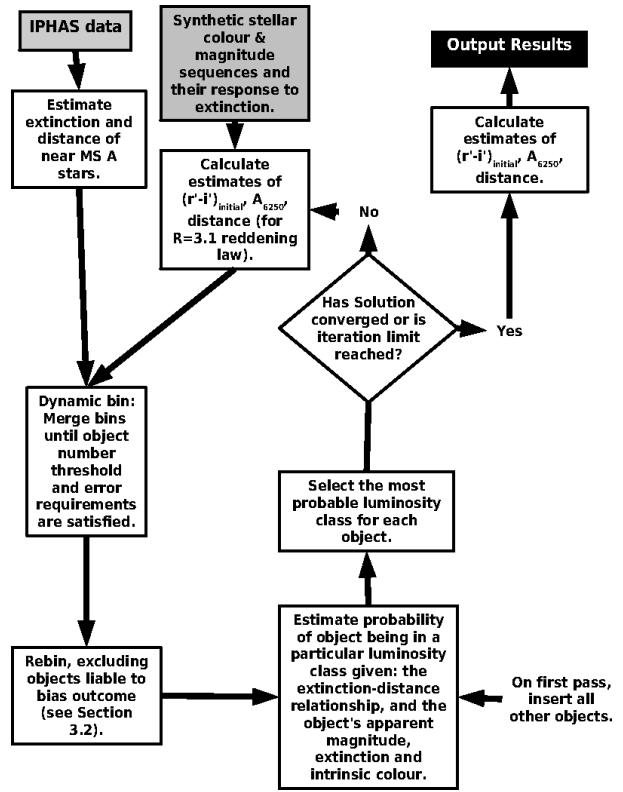}
\caption{A schematic depiction of {\scshape mead}.\label{alg_plan}}
\end{figure}

\subsection{Details of implementation}\label{Breakdown}

Fig.~\ref{alg_plan} outlines an algorithm that realises this principle (named {\scshape mead} for Mapper of Extinction Against Distance). In the remainder of this section each of the steps in {\scshape mead} are described.

Extinction, intrinsic colour and distance are determined simultaneously for all objects in the intrinsic colour range of interest given in Figs. \ref{HRD} and \ref{empiricalcompare}. Objects which fall outside this region of the colour-colour plane which could be occupied by objects of the types we are interested in are simply discarded. The discarded objects will include M and late-K type stars, white dwarfs, extragalactic objects and stars exhibiting H$\alpha$ emission. At the same time errors on the derived quantities are also  propagated from the photometric errors. Although determined at the same time as distance, the estimates of extinction and intrinsic colour are almost independent of the luminosity class and can be considered to be determined solely from the colour-colour plane: only very small differences arise from the slight differences between the luminosity class sequences in the colour-colour plane. 

The appearance of objects on the extinction-distance plane is complicated by several factors. Firstly, the size of the errors on the estimated values of extinction and distance vary greatly from object to object. Also, these errors are correlated with distance and extinction, as the apparent magnitude of a source becomes fainter with increasing distance or extinction, increased photometric error is to be expected. Finally, because of the finite angular resolution of the fields, there will inevitably be unresolved substructure within the absorbing material in each field analysed, causing an intrinsic and irreducible spread of extinctions at any given distance. These complications are dealt with by assuming that extinction at any given distance is modelled by a distribution other than the delta function: a maximum likelihood estimator is used to determine the value of the parameters that describe the distribution. The choice of family of distributions used here is further discussed in Section~\ref{coping_test}.

IPHAS observations are subject to both bright and faint magnitude limits. The implication of this is that for an object at a given distance, it will only be observed if its extinction also falls between two limits, and when these limits are approached, the observations will start to become incomplete. This would bias any estimates. The nature of the bias is that as the faint cut-off is approached, only stars viewed through lower extinction remain in the sample, while more extinguished objects are lost. \cite{Hakkila.1997} note the disruptive effect that this had on previous studies \citep{Fitzgerald.1968, Neckel.1980, Arenou.1992}.  Even at the higher angular resolutions treated here this effect continues to be a problem because the interstellar medium is inhomogeneous down to very small scales. \cite{Deshpande.2007} demonstrates that variations in the structure of the ISM naturally extend down to at least the scale of tens of AU. Therefore, it is necessary to continue to attend to the magnitude limits in this study, even though significantly finer angular resolutions are achievable than those in earlier work.

To deal with this problem, it is necessary to determine which objects could be close to the magnitude limit in distance extinction space, so that they can be excluded when producing an unbiased distance-extinction relationship. To establish which objects are to be excluded, first all the objects are binned in distance to produce a biased extinction-distance relationship, for each bin the mean and intrinsic standard deviation of the extinction is found. Then, for each object the maximum ($d_{\rm max}$) and minimum ($d_{\rm min}$) distances for which observations of that type of object would be complete are estimated. These distances occur where the following two equations are satisfied, where $A_{6250}^{\rm bright}(d)$ and $A_{6250}^{\rm faint}(d)$ correspond to the bright and faint magnitude limits respectively and $\sigma_{\bar A}(d)$ is the intrinsic standard deviation of extinction in the relevant bin:

\begin{equation}
\bar A_{6250}(d_{\rm max})+n\sigma_{\bar A}(d_{\rm max}) = A_{6250}^{\rm faint}(d).
\label{hi_limit}
\end{equation}

\begin{equation}
\bar A_{6250}(d_{\rm min})-n\sigma_{\bar A}(d_{\rm min}) = A_{6250}^{\rm bright}(d). 
\label{lo_limit}
\end{equation}

If these equations are satisfied at more than one distance, the largest possible value of $d_{\rm min}$ and smallest possible value of $d_{\rm max}$ are taken. In equations \ref{hi_limit} and \ref{lo_limit} the value of $n$ can be altered. Increasing it will reduce the chance of the bias affecting the data, at the cost of decreasing the number of objects which are binned to produce the distance-extinction relationship, whilst decreasing it will act in the opposite sense. There is no immediately obvious value to adopt for $n$ and Section~\ref{maglims_test} discusses the use of simulated photometry to find an appropriate value. At the bright end it is assumed that the magnitude range over which observations become incomplete is small, relative to the apparent magnitude sampling, so objects which are closer than $d_{\rm min}$ are simply discarded. In contrast, at the faint end where the apparent magnitude sampling is much denser, the magnitude range over which observations become incomplete is significant. Therefore the contribution of each object to the distance and extinction estimates is weighted using a sigmoid function as below:

\begin{equation}
\rm weight = \frac{1}{1+e^{s(\mu - \mu_{\rm max})}}.
\label{sigmoid}
\end{equation}

Where $\mu$ is the distance modulus of the object in question, $\mu_{\rm max}$ is the distance modulus corresponding to $d_{\rm max}$ and $s$ is related to the distance modulus range over which the sample becomes incomplete. A suitable value of $s$ was found empirically to be $3$. This was achieved by fitting a sigmoid to the faint end of histograms of the numbers of objects of a given spectral type against apparent magnitude. By applying this weight, objects which lie well beyond $d_{\rm max}$, where the sample is not complete for all expected extinctions, are given approximately zero weight, whilst those well within $d_{\rm max}$ are given a weight of almost one.

A Bayesian classification method is employed to estimate the luminosity class into which each object falls, given the determined distance-extinction relationship, the apparent magnitude of the object, an estimate of the extinction to the object and an estimate of its intrinsic colour. The luminosity class sequences are assumed to have Gaussian spread in absolute magnitude, with all sequences having equal and constant spread. The classification is made by evaluating the posterior probability that an object belongs to a luminosity class, given its extinction, the distance-extinction relationship and the absolute magnitude of that class. The choice of the prior used is discussed in Section~\ref{calibration_priors_test}. 

A threshold probability is imposed, such that when the probabilities of all five classes fall below this, the object is removed from consideration in that iteration. This removes objects which do not appear to fit into any of the luminosity classes. 

These procedures are put into an iterative structure, as shown in Fig.~\ref{alg_plan}. The sense of the iteration can be summarised as follows: the distance, extinction and intrinsic colour of each object is estimated; then these determinations are binned to get a distance-extinction relation; this is then used to determine the luminosity class of the objects; and then the distance, extinction and intrinsic colour are estimated again, given the new estimate of luminosity class. The process has shown good convergence in all cases examined so far. In the range of 5-10 iterations are normally required.  Convergence is discussed further in Section~\ref{basic_test}.

{\scshape mead} begins with the derivation of the distance-extinction relation defined by the early A-stars alone, that lie along the lower edge of the main stellar locus in the colour-colour plane.  This is used to obtain the first estimate of the luminosity classes of all objects in the field.  It is safe to assume that the great majority of objects on this strip are near main sequence, given that early-A giants are not only relatively rare but also exhibit slightly weaker $\Halpha$ absorption that lifts them to higher $(r'-\Halpha)$.  This breaks the degeneracy between the different luminosity classes and enables distances to be reliably estimated at the outset \citep{Drew.2008}.

When binning the data by distance {\scshape mead} ensures that: the depth of the bin is at least 100~pc; there are at least eight objects in the bin; the total signal to noise in the bin is at least 130. 

It should be noted that not all available data in the IPHAS database are used in this algorithm. Only data flagged as either stellar or probably stellar in all three bands by the processing pipeline are accepted, with the remainder discarded straight away \citep[for more details of the data processing see][]{Irwin.2001, Drew.2005short, Gonzalez-Solares.2008short}. As noted already in this section only a restricted stellar intrinsic colour range is exploited (marked in Figs. \ref{HRD} and \ref{empiricalcompare}): this is due in part to the \cite{Houk.1997} calibration ending at K4V and also to the difficulties associated with modelling the molecular bands in late type-stars (Section~\ref{calibration}). Following \cite{Drew.2008}, a maximum bright limit of $r'=13.5$, $i'=11.5$ and $\Halpha=12.5$ is employed to avoid objects that may be saturated in one or more bands. Occasionally when any fainter objects are classified as saturated by the pipeline, the limit will be moved to the apparent magnitude of that object.

\subsection{Possible contaminants}\label{contaminents_test}

{\scshape mead} assumes that all the objects examined can be described by one of the luminosity class sequences. In practice there are a minority of exotic objects that will not fit into them. Generally such objects are not a problem if they either: have colours which are outside the range {\scshape mead} looks at or are significantly less frequent than normal objects. The latter argument relies on the statistical nature of {\scshape mead}. Although the properties of individual objects may be incorrectly estimated, the intention is that {\scshape mead} will derive the correct extinction-distance relationship given a large enough sample.

As previously mentioned, O and B stars are contaminant objects that may be mistaken for A and F stars. Fortunately, O and B stars are relatively rare, and present only in young populations. In populations older than $\sim$100 Myrs they are essentially absent.  So, in large samples, with a mix of ages, we can comfortably assume field A and F stars dominate. The circumstance in which this assumption may come undone is in regions of ongoing star formation, especially if the stars that will become A and later-type main sequence stars are still in the pre-main-sequence phase. If such objects remain in the minority for their locale, MEAD should recognise their raised luminosity and classify accordingly.  If they become the dominant population, then MEAD may misinterpret them. In Section~\ref{6010} {\scshape mead} is tested against IPHAS observations of the Cygnus OB2 association, where such extreme conditions could exist.

The subdwarf sequence occupies a very similar position on the IPHAS colour-colour plane as its dwarf equivalent and as such there is the possibility that subdwarfs may be erroneously identified as dwarfs and used in {\scshape mead}. However, in a volume limited sample it is expected that roughly one in every $10^3$ objects would be a subdwarf \citep{Reid.2001}. Given the faint absolute magnitudes of subdwarfs relative to the dwarfs, this proportion becomes even lower in the magnitude limited IPHAS catalogue.  Furthermore, the imposition of a threshold probability in the Bayesian luminosity classifier will prompt {\scshape mead} to discard a large proportion of them as they are offset below the main sequence in colour-magnitude space.  

Hydrogen atmosphere white dwarfs largely lie outside the main stellar locus, leaving only a minority inside the region of the colour-colour plane searched by {\scshape mead}: these objects will be the very coolest and the very hottest white dwarfs. White dwarfs exhibit absolute magnitudes approximately in the range $10 \la r' \la 15$. Given this and the IPHAS magnitude limits, the database cannot contain even unreddened white dwarfs more distant than 1~kpc. It is estimated that there will be $7.6$ white dwarfs per square degree in the IPHAS sample, representing $\sim0.01\%$ of all sources. This estimate is based on a detailed model of the Galactic population of white dwarfs \citep{Napiwotzki.2008} calibrated using the observed white dwarf samples of \cite{Holberg.2008} and \cite{Pauli.2006}, reddening is included in an approximate way. Many white dwarfs will be ignored by {\scshape mead} due to their colours lying outside the main stellar locus.  Of the handful remaining within more will be discarded by the Bayesian luminosity classifier as belonging to a faint sequence. 

Extra-galactic sources will also contaminate the catalogue. Resolved galaxies will be flagged by the CASU pipeline as being extended objects and will be ignored by {\scshape mead}. In addition, even some small galaxy images will overlap with the very numerous stellar images and hence be classified  as extended, also removing them from consideration. Nevertheless, some galaxy images will be sufficiently compact to be classified as stellar. Genuinely compact (i.e. physically small) galaxies are rare, contributing about 3\% of all low redshift field galaxies down to $B \simeq 20$ ($r' \sim 19$) according to \cite{Drinkwater.1999}, who used the all-object (high latitude) spectroscopic survey of \cite{Drinkwater.2000}. Extrapolating this another magnitude deeper (i.e. to $r' = 20$), we would still only expect some 40 objects per square degree, which is completely negligible compared to (i.e. 0.02\% of) the 200000 stars per square degree in low latitude fields. Of course, at faint limits all galaxies tend to look less and less resolved, resulting in the well known merging of the stellar and galaxy sequences in star/galaxy separation diagrams \citep[e.g.][]{Driver.2003}. However, at $r' = 20$ there are around 2000 galaxies per square in total, approximately two thirds of which are in the final magnitude bin \citep[e.g.][]{Driver.1994}, so even if we assume that all these could look stellar, this would only represent a 0.7\% contamination of the ``stars''. In regions of high extinction, the counts would, of course, be reduced at given $r'$. For an asymptotic $A_r = 2$, for instance, the galaxy counts  would be reduced by a factor of about 8, though this would be compensated somewhat by the galaxies having smaller angular sizes as the low surface brightness outer parts are too faint to detect \citep*{Phillipps.1981}. This would reduce the contamination to $\sim 200$ per square degree or $\sim 0.1$\%. 

Quasars and other distant AGN will also show unresolved images, but again numbers are small. From the \cite{Drinkwater.2000} all-object survey, \cite{Meyer.2001} find 35 QSOs (of all colours) per square degree to $B \simeq 20.2$ ($r' \sim 19.5$). Given the steep quasar number counts \citep{Croom.2004}, this would be $\sim 70$ per square degree at $r' = 20$. \cite{Vanden_Berk.2005} find a somewhat lower number density in the $i'$ band and many QSOs may have colours too blue to be included in our sample here, so we can estimate a contamination rate of order 0.02\% (and less, of course in highly absorbed regions).

Finally we note there are several classes of emission line star, including symbiotic stars, cataclysmic variables and planetary nebulae, which lie above the main stellar locus on an IPHAS colour-colour plot and so are ignored by {\scshape mead} \citep{Corradi.2008}.

\section{Testing MEAD against simulated photometry}\label{simulated}

The reliability and precision of {\scshape mead} needs to be appraised, by testing its performance on sightlines where the distance-extinction relation is exactly known. As there are no such ideal sightlines, it is necessary to synthesise model sightlines resembling those encountered in the Galactic disc, to which photometric errors can be applied that are consistent with IPHAS data.  We do this using Monte-Carlo sampling techniques. These simulations also allow us to directly investigate the impact of factors such as metallicity on the performance of {\scshape mead}, in a way that would otherwise not be possible.

\begin{figure}
\centering
\includegraphics[width=80mm, height=60mm]{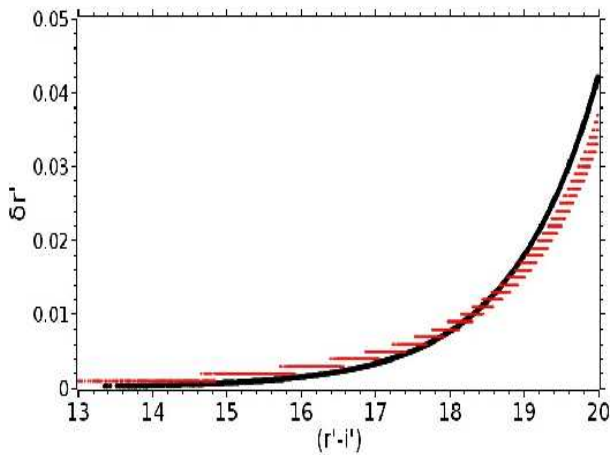}
\caption{The growth of random photometric errors in real data (red), from IPHAS field 2311o and simulated data (black), given similar seeing.\label{phot_error}}
\end{figure}

The simulations were created largely following the Besan\c{c}on galaxy model. The IMF of \cite{Kroupa.2001} and the galactic extinction model of \cite{Amores.2005} were assumed. The simulations were populated with objects from a copy of the Teramo library of isochrones and stellar evolutionary tracks \citep{Pietrinferni.2004}. These were translated into the IPHAS filter system using the finely-sampled grid of model SEDs due to \cite{Munari.2005}. Photometric errors were imposed, adopting random errors based on the CCD equation (see Fig.~\ref{phot_error}) along with field wide systematic errors of order 1.5\%, arising from the calculation of the aperture correction and photometric zero-points.

\subsection{Comparisons between simulations and real data}

Although it would be possible to create an end to end simulation, whereby synthetic images are created and then pipelined in the same manner as IPHAS observations, this would be so excessively time consuming as to restrict the analysis of {\scshape mead}'s performance. Instead, the simpler method used here begins with the application of {\scshape mead} to the synthesised magnitudes.  This allows the production of many synthetic lines of sight, which in turn enables {\scshape mead} to be robustly analysed. 

\begin{figure*}
\centering
\includegraphics[width=160mm, height=120mm]{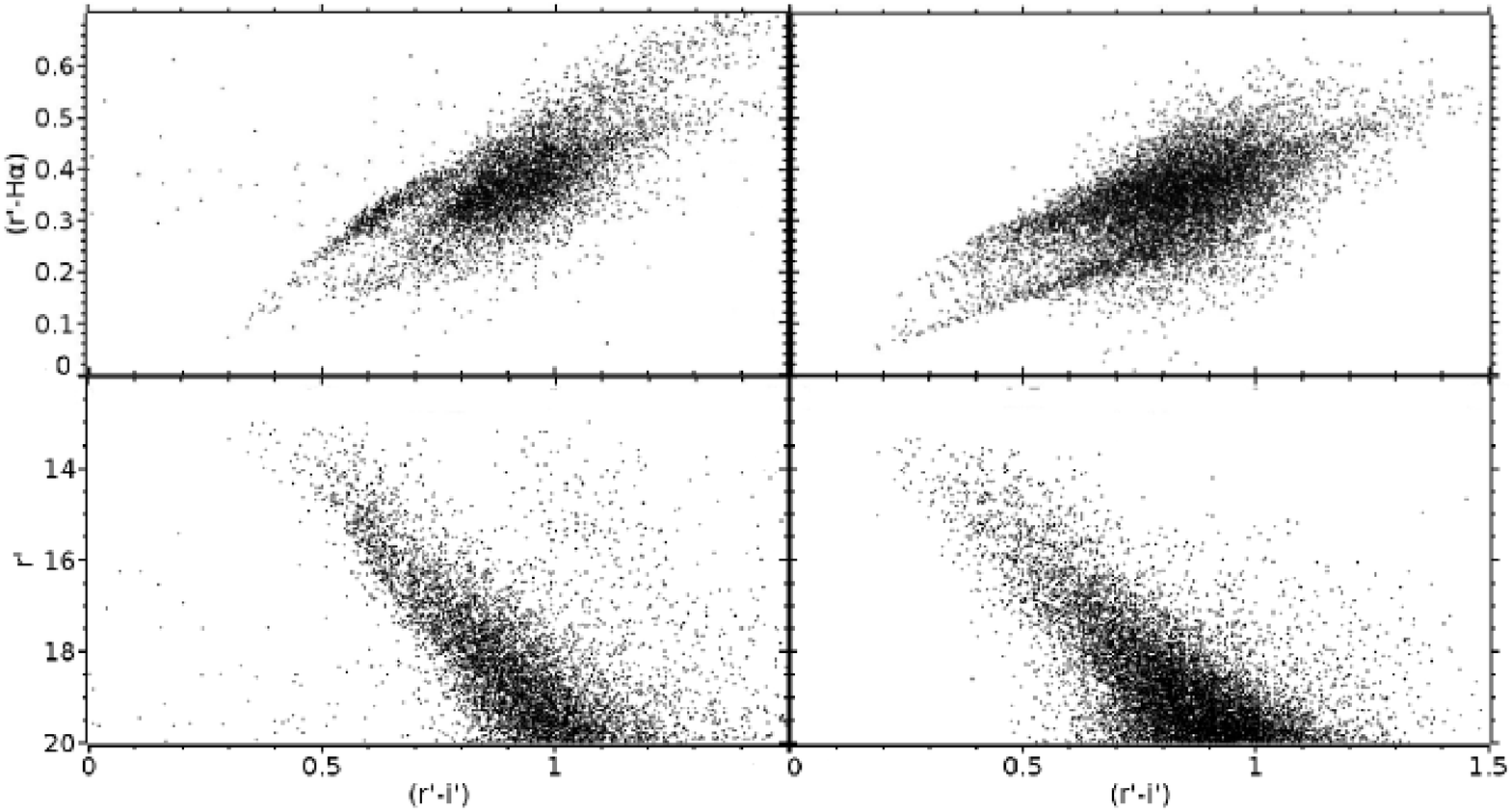}
\caption{A comparison between observed (left) and simulated (right), colour-colour (top) and colour-magnitude (bottom) diagrams , given similar conditions, in the direction $(l,b)=(170,0)$. The observed data are the observations of IPHAS field 2311o and include only objects classified as stellar or probably stellar. A magnitude cut at $r'=20$ has been applied to both the observed and real data. The differences between the two sets of data are due to: differences between the real and model distance-extinction relationship; differences between the distributions of objects with respect to distance; and some types of objects not being included in the simulations. \label{all_compare}}
\end{figure*}

The colour-colour and colour-magnitude plots produced by the simulations bear strong similarity to the observed plots, as shown in Fig.~\ref{all_compare}, though there are a few key differences. Significantly, the clear group of M-dwarfs are not reproduced in the simulations, due to the lower mass limit of the Teramo library. This is not a problem in this context, as {\scshape mead} does not use objects which occupy the region of the colour-colour plot occupied by M-type stars. The same argument applies to the exotic object types discussed earlier (see Section~\ref{contaminents_test}). These are not included in the simulations, as they do not significantly influence {\scshape mead}.

The difference between the two plots portrays the divergence that is to be expected between Galaxy models and the details of the real Galaxy. The \cite{Amores.2005} and real extinction-distance relations are in clear disagreement, giving rise to the different angles made by the main stellar loci with respect to the axes in the colour-magnitude diagrams of Fig.~\ref{all_compare}. The real colour-colour diagram in Fig.~\ref{all_compare} shows two principal groupings, one with low reddening, as seen in the group that runs through $(r'-i', r'-\Halpha) \sim (0.6, 0.3)$, while a second group with higher reddening occurs around $\sim (0.9, 0.35)$. These two groupings occur because the stars in this direction are not distributed smoothly with extinction, but rather are either local unreddened stars or are more heavily reddened and therefore probably more distant. The discontinuity may arise either as a result of the stars clumping together into two groups at different distances, or as a result of a sudden increase in extinction. The simulations, on the other hand, do assume that stars are distributed smoothly and extinction grows steadily, so there is no grouping visible.

\subsection{The basic performance of MEAD}\label{basic_test}

\begin{figure}
\centering
\includegraphics[width=80mm, height=60mm]{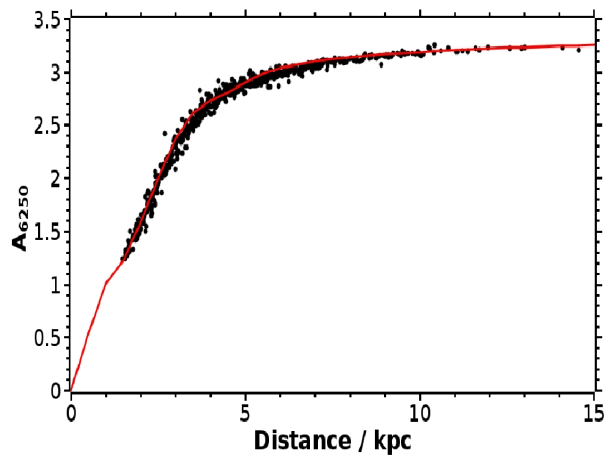}
\caption{The determined points on the extinction-distance relationship from twenty five simulations (black) and the underlying extinction, input to the simulations (red). \label{basic}}
\end{figure}

\begin{figure}
\centering
\includegraphics[width=80mm, height=60mm]{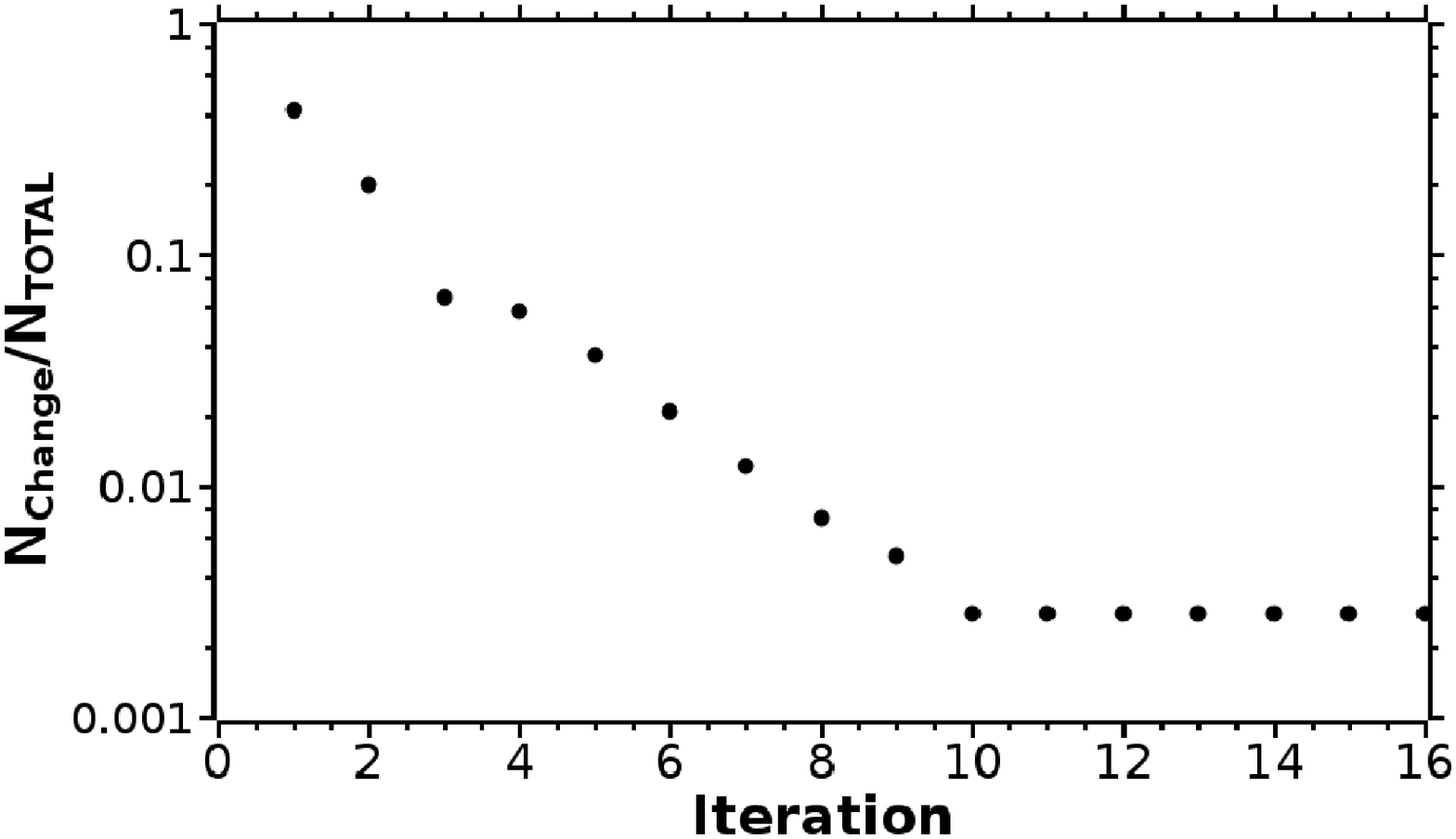}
\caption{Demonstrating the convergence of {\scshape mead} by showing how the number of objects which have their estimated luminosity class changed varies with iteration. From iteration 10 onwards {\scshape mead} is oscillating between two solutions, in doing so the luminosity class estimates of four objects are changed in each iteration. \label{converge1}}
\end{figure}

\begin{figure}
\centering
\includegraphics[width=80mm, height=60mm]{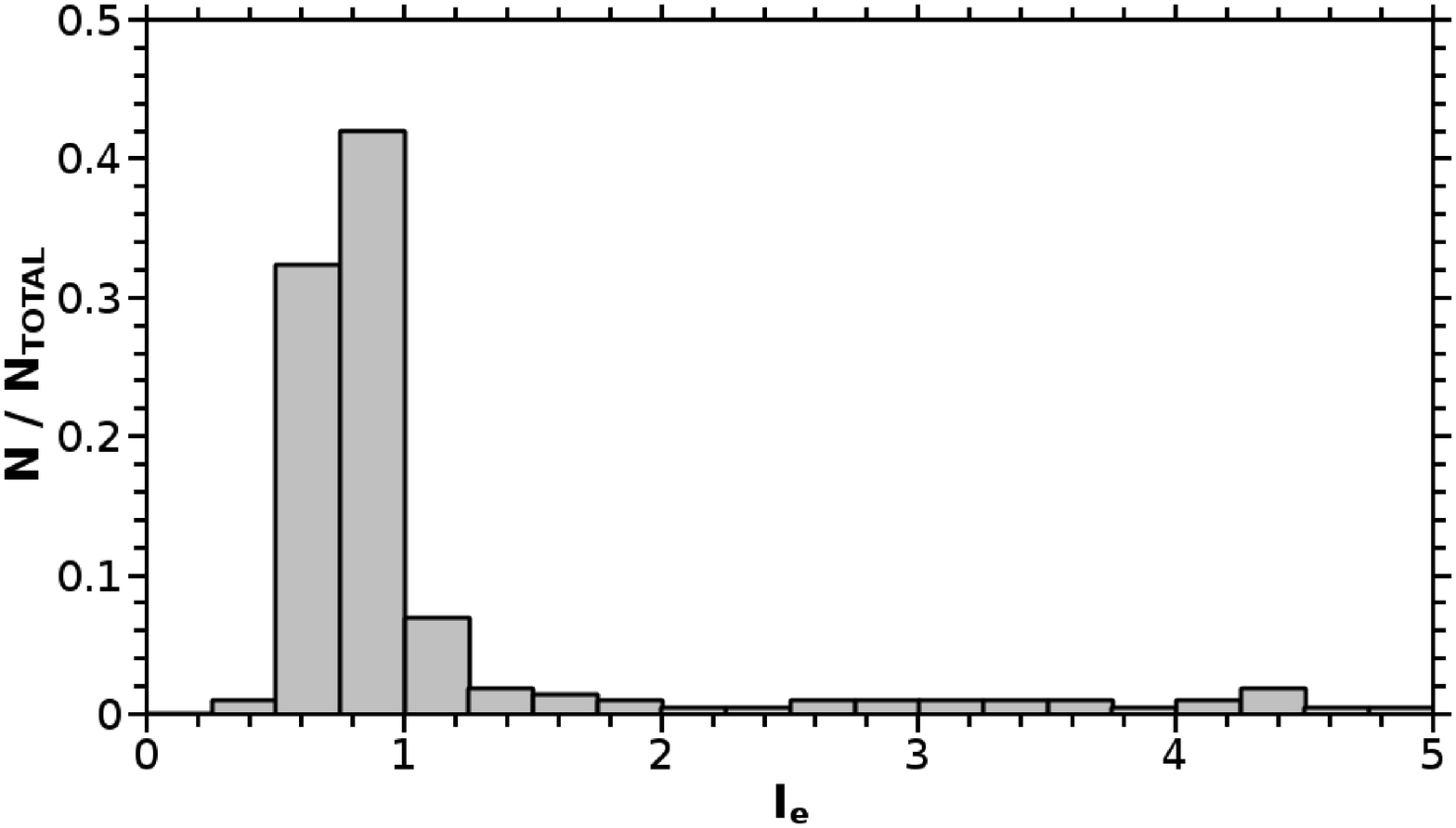}
\caption{ I$_{e}$ is the number of iterations required for the number of objects which have their estimate of luminosity class altered to drop by a factor of $e$, as determined by linear regression. I$_e$ is a measure of the gradient of plots like Fig.~\ref{converge1}, where lower values of I$_e$ correspond to steeper gradients and so quicker convergence.\label{converge2}}
\end{figure}

With the photometric errors, metallicity variation and binary fraction set to zero, {\scshape mead} is very successful at retrieving the input distance-extinction relation, as shown in Fig.~\ref{basic}. The little spread on extinction that does exist is a result of describing the continuous stellar evolutionary tracks in terms of 5 discrete luminosity classes (see Fig.~\ref{HRD}) . In this manner we can see that the errors induced by this description are small ($\delta A_{r'} \sim 0.05$) and so this is a valid method.

To be credible, the solution must converge.  Although {\scshape mead} will exit when complete convergence has been reached it is unreasonable to wait for every last small change to be made, especially if the last corrections to a few stars become oscillatory.  Fig.~\ref{converge1} demonstrates the convergence rate for one execution of {\scshape mead}, whilst Fig.~\ref{converge2} characterises the convergence rate for a large number of visualisations. The default maximum number of iterations for {\scshape mead} is set to 15.

\subsection{Calibration of priors for the determination of luminosity class}\label{calibration_priors_test}

In Section~\ref{algorithm} it was noted that it was possible to include a non-uniform prior probability when determining the luminosity class of a given object. Using the simulated photometry the probability distribution for the prior which retrieves the most accurate result was determined from a representative sample of sightlines. The accuracy is defined in terms of minimising the $\chi^2$ statistic: 
\begin{equation}
\chi^2=\sum\limits_{i} \frac{(A_{est}-A_{act})^2}{\sigma_{\bar{A}}^2+\delta A_{i}^{2}},
\label{chi_stat}
\end{equation}
where $A_{est}$ is the extinction estimated by {\scshape mead} and $A_{act}$ the extinction from the model at the same distance.  In the denominator, $\sigma_{\bar{A}}$ is the intrinsic standard deviation of extinction in the distance bin (i.e. due to the porosity of the ISM), while $\delta A_{i}$ is the error in $A_{est}$ as a consequence of the photometric errors.

Given that simulating sufficient sightlines is relatively time-consuming, it was not possible to search the entire parameter space, whilst the uncertain structure of the parameter space made 'downhill-only' root finding algorithms inadvisable. Therefore, a simulated annealing algorithm was used to minimise the statistic above. It was found that the prior should take the following form, where $P(X)$ is the probability assigned to luminosity class X: $P(V)=0.34; P(IV)=0.31; P(III)=0.28; P(II)=0.06; P(I)=0.01;$. This prior mirrors the number of objects in each class in the simulated photometry. As the IPHAS bright magnitude limit is at $r' \sim 13.5$, neither bright-giants ($M_{r'} \la -2$) nor super-giants ($M_{r'} \la -6.5$) are likely to be present in large numbers.

We note that this set of priors has been derived from a weighting of sightlines that, at the present time, emphasises the moderately reddened directions in the Galactic Plane encountered at all latitudes outside the Solar Circle, and at latitudes $|b| > 1$ within it. This set was specifically chosen to correspond to those sightlines investigated in this work.  For these sightlines dwarfs are strongly dominant and so the results show little sensitivity to the choice of priors. The priors for the highest reddening sightlines passing into the inner Galaxy are under investigation. It may prove necessary to attribute somewhat increased probabilities to brighter luminosity classes.   

\subsection{Coping with unresolved substructure \& photometric errors}\label{coping_test}

In {\scshape mead}, once estimates of distance and extinction are produced for each object, the data are binned, with respect to distance, to produce a distance-extinction relationship. As discussed in Section~\ref{algorithm}, the extinction will not be uniform for a finite sized box at any given distance. Instead it will vary with sky angle, just as it will vary with depth, due to unresolved substructure prompting some intrinsic variation in extinction. Hence, in this study, the extinction in any given bin is characterised as having some intrinsic scatter.

If the photometric errors are propagated through to give the error on the determined value of the extinction, it is possible to determine Maximum Likelihood Estimators (MLEs) for the mean and intrinsic standard deviation of the extinction, if the character of the variation in extinction is known. If the variation of extinction is Gaussian, then the estimate of the mean value of the extinction in each bin ($\bar A $) is given by the following equation: 
\begin{equation}
\bar{A}=\frac{\sum\limits_{i} \frac{A_{i}}{\sigma_{\bar{A}}^2+\delta A_{i}^2}}{\sum\limits_{i} \frac{1}{\sigma_{\bar{A}}^2+\delta A_{i}^2}},
\end{equation}
where $A_{i}$ is the estimated extinction for a single object, and other terms are as described in Section~\ref{calibration_priors_test}:

This estimate is essentially a weighted mean of the individual estimates of extinctions for each object in the bin. It is found that the intrinsic scatter of extinction in each bin ($\sigma_{\bar{A}}$) can be satisfactorily estimated by the weighted standard deviation of the extinctions of the brighter objects in the bin (those with $dA_{6250} \leq 0.1$). The weighting for this estimate is the weight given by equation~\ref{sigmoid} over $dA_{6250}^2$.

\begin{figure}
\centering
\includegraphics[width=80mm, height=60mm]{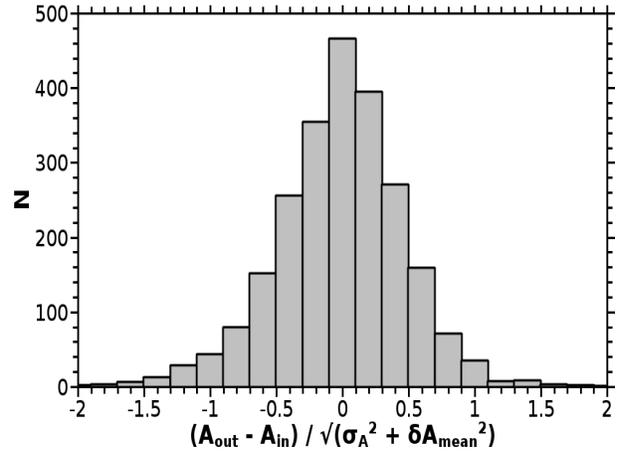}
\caption{The difference between the values of mean monochromatic extinction at $6250 \rm{\AA}$ at a given distance determined by {\scshape mead} ($A_{out}$) and the value in the simulation ($A_{in}$). The simulation has had intrinsic scatter in extinction imposed in a manner consistent with Kolmogorov turbulence in the ISM. \label{proper_scatter}}
\end{figure}

In reality, the intrinsic scatter of extinction in each distance bin is a result of the turbulent processes at work in the ISM. If the extinction of each simulated object is scattered away from the mean value for the distance as a result of Kolmogorov turbulence, {\scshape mead} still successfully retrieves the input extinction, despite the in-built assumption that extinction at a given distance is scattered normally (Fig.~\ref{proper_scatter}). 

\subsection{Handling of magnitude limits}\label{maglims_test}

\begin{figure}
\centering
\includegraphics[width=80mm, height=110mm]{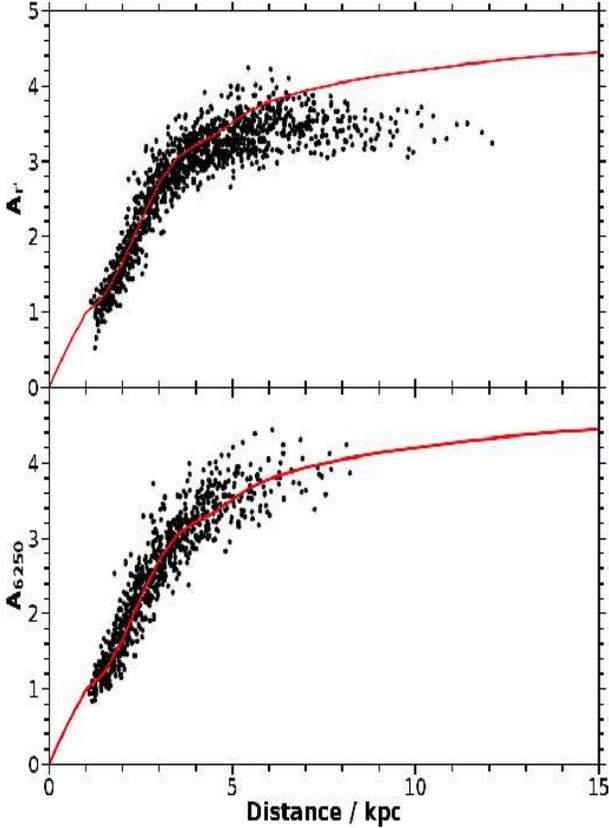}
\caption{Setting the value of $n$ too low in equations \ref{hi_limit} and \ref{lo_limit}, thus compromising the results of {\scshape mead}. Here the upper panel shows $n=0.5$ and  the lower $n=2.0$. Both panels are for 50 iterations in the direction $(l,b)=(180,0)$, the actual (input) distance-extinction relation is the red line and the black points are the results from {\scshape mead}. \label{lowx_problem}}
\end{figure}

\begin{figure}
\centering
\includegraphics[width=80mm, height=60mm]{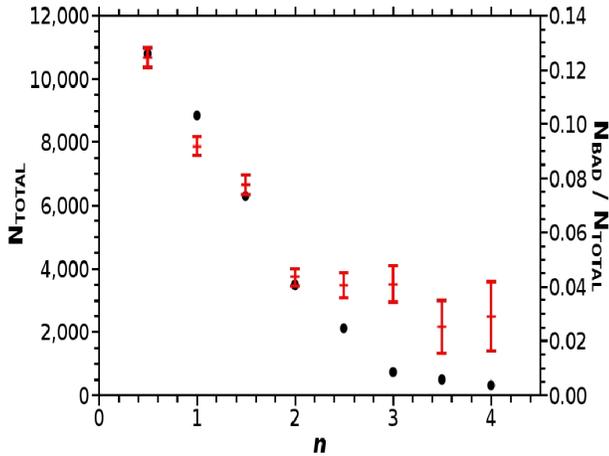}
\caption{The variation in the total number of data points (black dots, left axis) and proportion of bad data points (red crosses with error bars, right axis) on the derived distance-extinction relation as a function of $n$ for the directions $(l,b)=(80,0), (130,0), (180,0)$. Where good bins are those which satisfy the condition $A_{out}-A_{in}<\sigma_{A}^2+\delta_{\bar{A}}^2$ and bad bins are those which do not.\label{x_select}}
\end{figure}

In Section~\ref{Breakdown} we discuss the difficulties associated with the magnitude limits and how these have affected previous extinction maps. The treatment for this problem involves ignoring data which are at a distance where they could be affected by incompleteness. The parameter $n$ in equations \ref{hi_limit} and \ref{lo_limit} determines which data are ignored, but there is no obvious value it should take. To establish the ideal value of $n$, {\scshape mead} was tested against variation of this parameter. Setting the value of $n$ too low allows the inclusion of objects which may bias the results, due to the effect of incompleteness becoming increasingly relevant. It was found that for $n<2$ the results of {\scshape mead} would be unsatisfactory, an example of this effect is shown in Fig.~\ref{lowx_problem}. Raising $n$ acts to reduce the number of objects which can be used to determine the distance-extinction relationship, thus decreasing the range and distance resolution of the outcome. Although many of the objects filtered out should be removed for reasons of incompleteness, others need not be and are lost undeservedly. A compromise is necessary. We set $n=2$ so as to use the maximum possible number of objects whilst avoiding the magnitude limit effect. Fig.~\ref{x_select} illustrates that setting $n=2$ obtains the highest number of accurate distance bins on the derived distance-extinction relationship whilst maintaining a low proportion of bad bins.

\subsection{Achievable angular resolution}\label{angular_res_test}

Previous observational 3D extinction maps of the Milky Way \citep[e.g.]{Fitzgerald.1968, Lucke.1978, Neckel.1980, Pandey.1987, Joshi.2005} have been carried out at angular resolutions of the order of a few degrees. The recent extinction map due to \cite{Marshall.2006} has a resolution of $30 \arcmin$. With the high angular density of sources in the IPHAS database ($\sim 10^5 sq.deg.^{-1}$) it is clearly possible to map extinction at a much finer angular resolution. The resolutions possible are dependent on the source density, but even in the relatively sparsely populated galactic anti-centre resolutions of $10 \arcmin$ are achievable.

\subsection{Bias in distance estimation}

The estimate of the distance modulus to each object has a symmetric error distribution around its mean. However, the expected distribution of stars is not symmetric as the volume of space in which a star lies and the density of stars changes. Therefore, the expectation of the true distance to the star does not equal the original estimate and so that original estimate is biased. An analogous effect leads to the well known Lutz-Kelker bias \citep{Lutz.1973}, when using trigonometric parallaxes to estimate the absolute magnitudes of objects. Following \cite{Lutz.1973} it is possible to derive a correction for this bias, if the density of stars is assumed to be uniform. If this correction is applied to the distance bins of objects from the simulated photometry, where the stellar density is assumed to be uniform, the median correction is found to be $0.1\%$, whilst $95\%$ of bins have corrections less than $1\%$. The improvement on the agreement between the input and derived extinction-distance relationships is thus negligible.

Calculating a correction for a realistic density distribution is considerably more difficult. However, the median correction will often be smaller than in the case of uniform stellar density, because of the drop off in stellar density associated with most sightlines.

\subsection{Sensitivity to disc metallicity gradients}\label{metal_test}

The Besan\c{c}on model assumes a smooth gradient in the mean metallicity with respect to galactocentric radius ($R_G$). This view is shared by \cite{Friel.2002} and \cite{Bragaglia.2006} amongst others. However, it has been argued \citep*[e.g.][]{Twarog.1997, Corder.2001, Yong.2005} that metallicity in fact varies as a step function of $R_G$, with the step occurring at $R_G \sim10$~kpc. \cite{Carraro.2007} and \cite{Bragaglia.2008} argue that any gradient flattens to a constant metallicity in the outer disc. Given this spread in opinion it was necessary to test the performance of {\scshape mead} against a range of different schemes.

It was previously shown in Fig.~\ref{metal_cc} that metallicity has almost no effect on the colours of A0-K4 stars. The absolute magnitudes, on the other hand, are metallicity dependent, but these do not become significant until metallicities are markedly sub-solar (Fig.~\ref{metal_cm}).  Consistent with this we find that introducing metallicity variations induces little change in the output from {\scshape mead}, even though {\scshape mead} assumes solar metallicity. This is true of all the above mentioned schemes for the disc metallicity gradient.

\subsection{Correction for binarity}\label{binarity_test}

Unresolved binaries and higher multiples present the possibility of the systematic over-estimation of point-source absolute magnitudes. For the objects selected by {\scshape mead} (A0-K4), the effect of binarity on the observed colours is small \citep{Hurley.1998}. 

In the extreme case of binaries of unit mass ratio the distance to an object would be underestimated by a factor of $\sqrt{2}$ compared to the single-star case. However, on making plausible allowances for the binary mass ratio distribution and for the binary fraction, it is likely that, in the mean, distances will only be underestimated by a factor of a few percent. For example, the expected underestimate of distance will be $\sim 5\%$ if the \cite{Kroupa.2001} IMF, \cite{Duquennoy.1991} binary fraction, a constant probability function for binary mass ratio and a mass-luminosity relation of $L \propto M^{3.5}$ are all assumed. 

To arrive at a correction factor that may be applied routinely within {\scshape mead}, the effect of binarity has been analysed through the use of simulated photometry. Binaries were inserted in the model using a binary fraction of $57\%$, following \cite{Duquennoy.1991} which is derived from observations of G-type stars. \cite{Lada.2006} demonstrates that the binary fraction drops significantly for later type stars, but these are mostly those which are not simulated (i.e. M-type stars and later). The probability function for the binary mass ratio was assumed to be constant, whilst higher order unresolved multiples are assumed to be relatively rare \citep{Duquennoy.1991} and so are not included in the simulation.

The outcome of this exercise is that there is systematic underestimation of distances, which in turn causes extinction to be overestimated in a given distance bin. The appropriate correction for this is made by multiplying all the derived distances by $1.06$. This figure was derived through minimisation of the statistic in equation~\ref{chi_stat}, over 100 visualisations of 6 different sightlines. This modification should function successfully as most bins contain relatively large numbers of objects and the correction itself is small. When this correction is adopted {\scshape mead} successfully reproduces the input distance-extinction relationship.

\subsection{Impact of reddening law variations}\label{law_test}

\begin{figure}
\centering
\includegraphics[width=80mm, height=60mm]{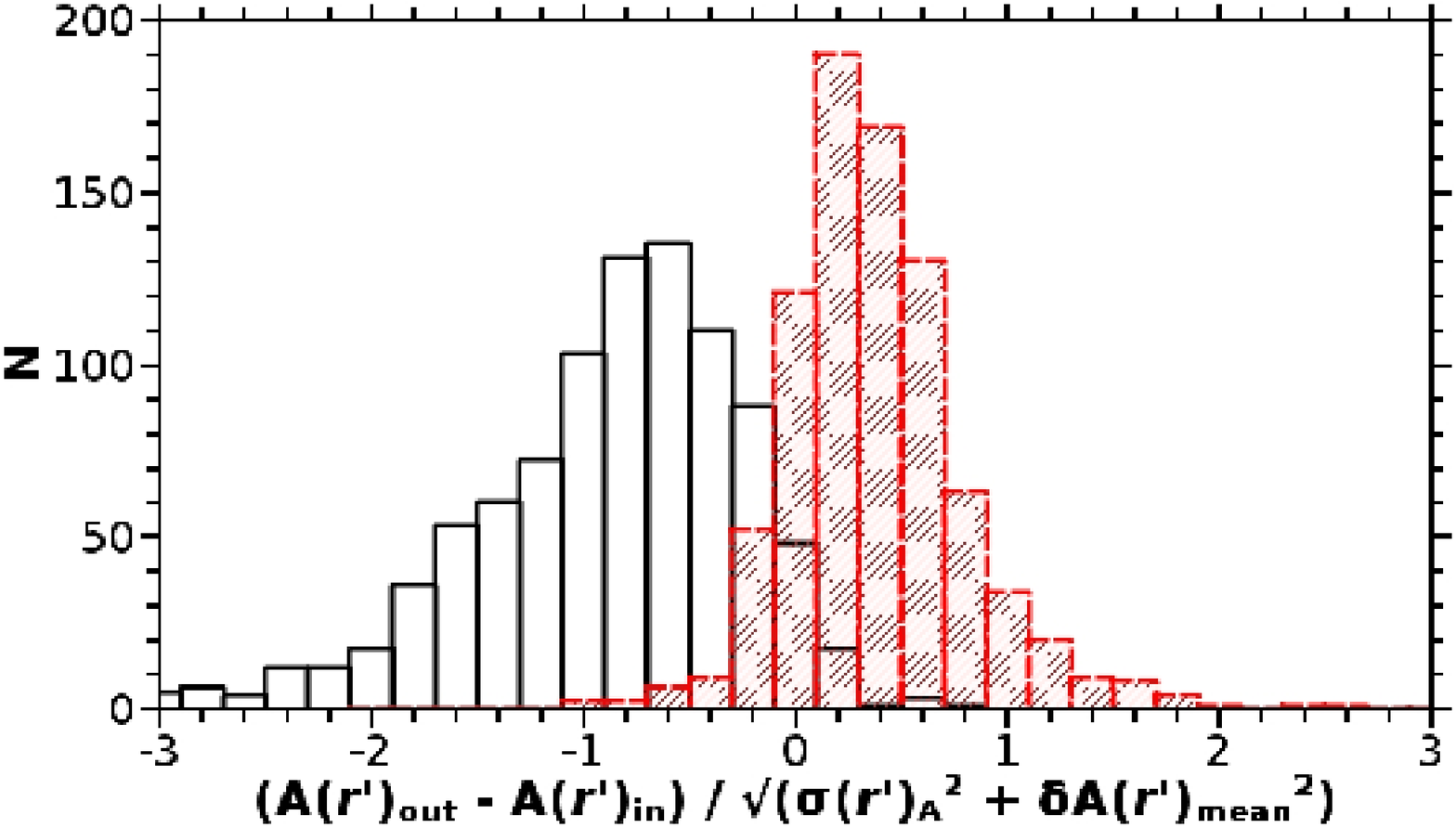}
\caption{The effect of altering the value of $R$ on the difference between the values of mean broadband extinction in the $r'$ band at a given distance determined by {\scshape mead} ($A(r')_{out}$) and the value in the simulation ($A(r')_{in}$). The results for $R=3.6$ are represented by unfilled black bars bounded by solid lines, $R=2.6$ by hatched red bars bounded by dashed lines. The simulations are for the direction $(l,b)=(180,0)$ and 50 visualisations were performed in each case. \label{R_test_r}}
\end{figure}

{\scshape mead} and the simulations performed up until this point assume the $R=3.1$ reddening law of \cite{Fitzpatrick.2004}. To investigate the effect of altering $R$ significantly, simulated photometry was created with $R=2.6$ and $3.6$ reddening laws. Then {\scshape mead} was run on the results of these simulations, while continuing to apply the $R=3.1$ law. Fig.~\ref{R_test_r} shows that the impact on {\scshape mead}'s mapping of sightlines better represented by non-standard reddening laws is to induce a systematic error that is comparable - in these examples - with the typical random error.  Had the reddening been expressed in terms of the monochromatic extinction, $A_{6250}$, Fig.~\ref{R_test_r} would look much the same.  However, a conversion into $A_{V}$, would be even more dependent on the value of $R$ and therefore the outcome of a mapping expressed in terms of $A_V$ would be subject to even greater systematic error.  The sense of the bias is that the extinction at a given distance for sightlines of anomalously high $R$ is underestimated.

Given that IPHAS observations provide magnitudes in just three bands, it is not feasible to constrain $R$ in addition to the intrinsic colour, extinction and luminosity class (and so distance). Instead it is preferable to rely on the relative dominance of the $R=3.1$ reddening law across the Galactic Plane. The same assumption has been made previously by \cite{Neckel.1980, Schlegel.1998, Marshall.2006} and many others. If the IPHAS database were to be cross-matched with the results of the United Kingdom Infrared Deep Sky Survey Galactic Plane Survey \cite[UKIDSS GPS;][]{Lucas.2007short} and the Ultraviolet Excess Survey of the northern Galactic Plane (UVEX; Groot et al. in preparation) the extra measurements would facilitate the accurate estimation of $R$, whilst still maintaining the accuracy of the other estimates.

\subsection{The impact of very young populations}\label{young_pops_test}

O, B and very early A stars occupy the same area of the IPHAS colour-colour plane as less-extinguished later A and F stars. However, they also respond differently to extinction and have different absolute magnitudes and so introduce a degeneracy into the problem. In the inter-arm field such objects are comparatively rare, with their short lifespans and the IMF suppressing their frequency. However, in the spiral arms such objects are more frequent, due to the on-going star formation induced by the passage of the spiral density wave. The Besan\c{c}on model makes no allowance for spiral-arm induced star formation, since a constant star formation rate is adopted for all locations. Therefore simulations based on the Galactic model cannot provide any insight into the impact of these localised young populations. We have carried out some tests of such concentrations by inserting them into model sightlines in order to mimic the superposition of a spiral arm.  The results show no additional systematic error in the derived extinction curves that would imply a breakdown of the method.

As with early-type stars, pre-main sequence (PMS) objects introduce degeneracy, due to their different colours, absolute magnitudes and response to extinction as compared to the sequences assumed in the calibration of {\scshape mead}. Also, as with early-type stars, PMS objects would normally be found concentrated in star forming regions. So as to investigate the effect of including PMS objects the extra spiral arm population was again included in the model. The inserted PMS objects were then tracked through {\scshape mead}. In reality, many PMS objects exhibit $\Halpha$ emission and so lie outside the main stellar locus on colour-colour plots and are thus ignored by {\scshape mead}. However, those PMS objects exhibiting weaker $\Halpha$ emission would remain inside the main stellar locus, where they may be misidentified by {\scshape mead}. As with O and B stars, the relative scarcity of emission-line PMS stars compared to normal stars should limit this problem. As a direct test of this and other expectations regarding young populations, we present and discuss the example of sightlines through Cyg~OB2, a spectacular northern OB association, in section~\ref{6010}.

\subsection{The impact of photometric zero-point errors}\label{need}

At the time of writing the IPHAS database lacks a global calibration, with observations instead being calibrated on a run by run basis, with the $\Halpha$ calibration tied to the $r'$ calibration. As such it is possible that there will be offsets between the true and measured magnitudes in one or more bands.  To investigate the effects of such difficulties, offsets were manually introduced into the simulated photometry. 

The effect of adding the same magnitude offset to all three bands is simple to understand, as this will simply lead to inaccurate determinations of the distance to each object. For example an offset of $0.10$ mags is well within the range of what might be encountered within the IPHAS database presently.  This offset leads to a $5\%$ error in determined distances.

\begin{figure}
\centering
\includegraphics[width=80mm, height=110mm]{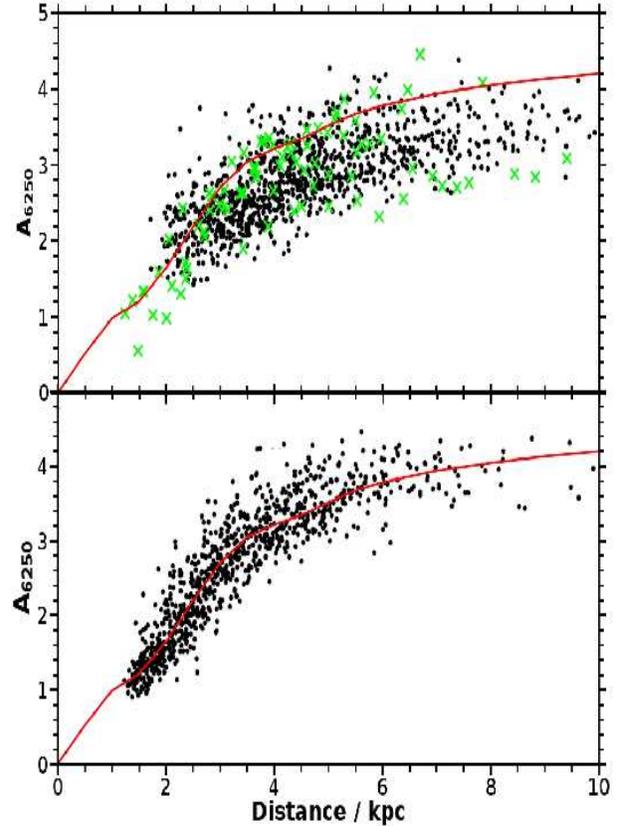}
\caption{The top panel shows the retrieved extinction-distance relationship, given an offset of $0.05$ (black dots) or $-0.05$ (green crosses) to the $\Halpha$ magnitudes of the objects. The bottom panel shows the results with no offset (black dots). On both panels the input relationship is shown as the red line. The simulations are for the direction $(l,b)=(180,0)$. \label{HA_off}}
\end{figure}

Somewhat less straightforward is the effect of offsetting one band only, as this will affect the intrinsic colours and extinctions determined for each object, and so the object distance as well. Fig.~\ref{HA_off} shows the effect of applying a positive and negative offset to $\Halpha$. There are two consequences of doing this. The first is that the accuracy of {\scshape mead} is reduced, as all objects are now systematically misidentified. This effect can be seen clearly by comparing the two panels in Fig.~\ref{HA_off}. Second, {\scshape mead} returns fewer points, for the case of a negative offset to $\Halpha$.  This can be very serious as it may reduce the number of returned points significantly.  In the case shown $90\%$ of the objects available are lost: this is because the main stellar locus has shifted largely outside the region of the colour-colour diagram searched by {\scshape mead}. 

Until a global photometric calibration is performed, tying the $\Halpha$ calibration to $r'$ vastly reduces the frequency and size of offsets in the ($r' - \Halpha$). In the future the calibration in the $i'$ band will also be tied to the $r'$ band calibration, improving the suitability of the data for use in {\scshape mead}.

\section{Example applications of MEAD}\label{examples}

\subsection{A sightline in Aquila, through ($l,b = 32.0, 2.0$)}

\begin{figure}
\centering
\includegraphics[width=80mm, height=60mm]{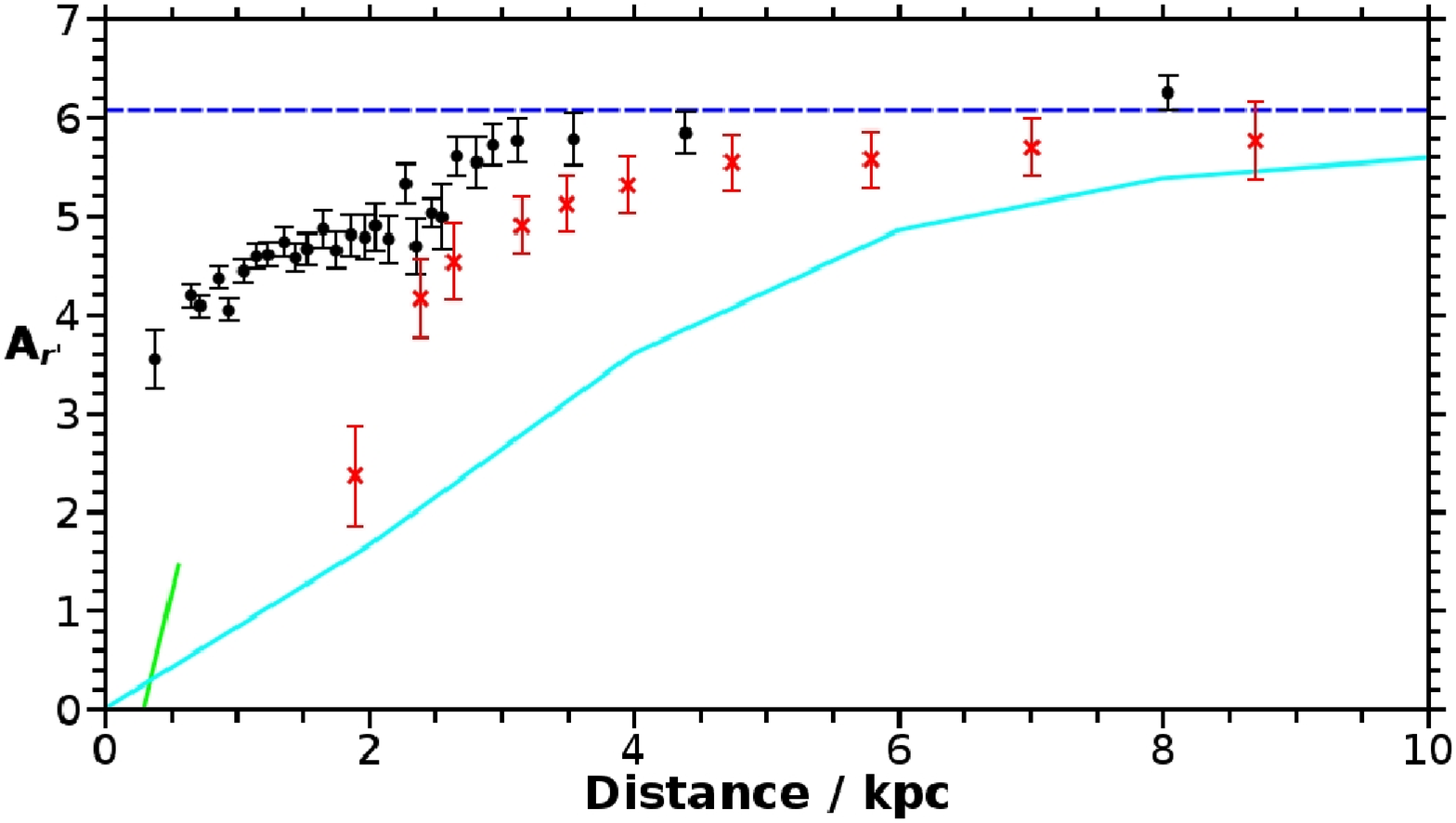}
\caption{Distance-extinction relationship for field 4199 in the direction $(l,b)=(32.0, 2.0)$, using a $10 \arcmin \times 10 \arcmin$ box. The relationship derived in this work is in black, with the error bars indicated the errors on the derived mean value of extinction in each bin. The result of \protect\cite{Neckel.1980} is in green, the \protect\cite{Schlegel.1998} asymptotic value is shown with the blue dashed line, the \protect\cite{Drimmel.2001} model in cyan and the \protect\cite{Marshall.2006} results are shown with red crosses. Estimates of extinction and distance for 1197 objects have been binned up to produce the distance-extinction relationship.\label{4199}}
\end{figure}

For this direction, the results from this work (Fig.~\ref{4199}) show two significant increases of extinction, one within the first few hundred parsecs, which would be associated with the Aquila Rift. The Rift is at a distance of $\sim 200$~pc \citep{Dame.1987}, with the extinction rising to $A_{V} \simeq 3$ ($A_{r'} \simeq 2.5$) \citep*{Straizys.2003}.  We find that the extinction through the Rift is somewhat higher, reaching $A_V \sim 4.7$ (or $A_{r'} \sim 4$) at this position. The second extinction rise lies at $\sim2$~kpc and is likely to be associated with the Sagittarius Arm. The extinction then asymptotes beyond it at a value roughly consistent with the \cite{Schlegel.1998} limit for this direction.

Previous extinction-distance relations for this sightline are also shown in Fig.~\ref{4199}.  At distances greater than 2~kpc, {\scshape mead} produces a relation that is quite similar to that of \cite{Marshall.2006}.  It is clear, however, that the \cite{Marshall.2006} modelling does not take into account the more local Aquila Rift.  It is this difference, in particular, that lies behind the great discrepancy between their first data point at $\sim$2~kpc and our relation.  At all distances we find the extinction to be markedly higher than given by the relation due to \cite{Drimmel.2001}.  These distinctions begin to illustrate the value of an approach that does not depend on Galactic models.

\subsection{The sightline through the open cluster NGC 2099 ($l,b = 177.6, 3.1$)}

\begin{figure}
\centering
\includegraphics[width=80mm, height=60mm]{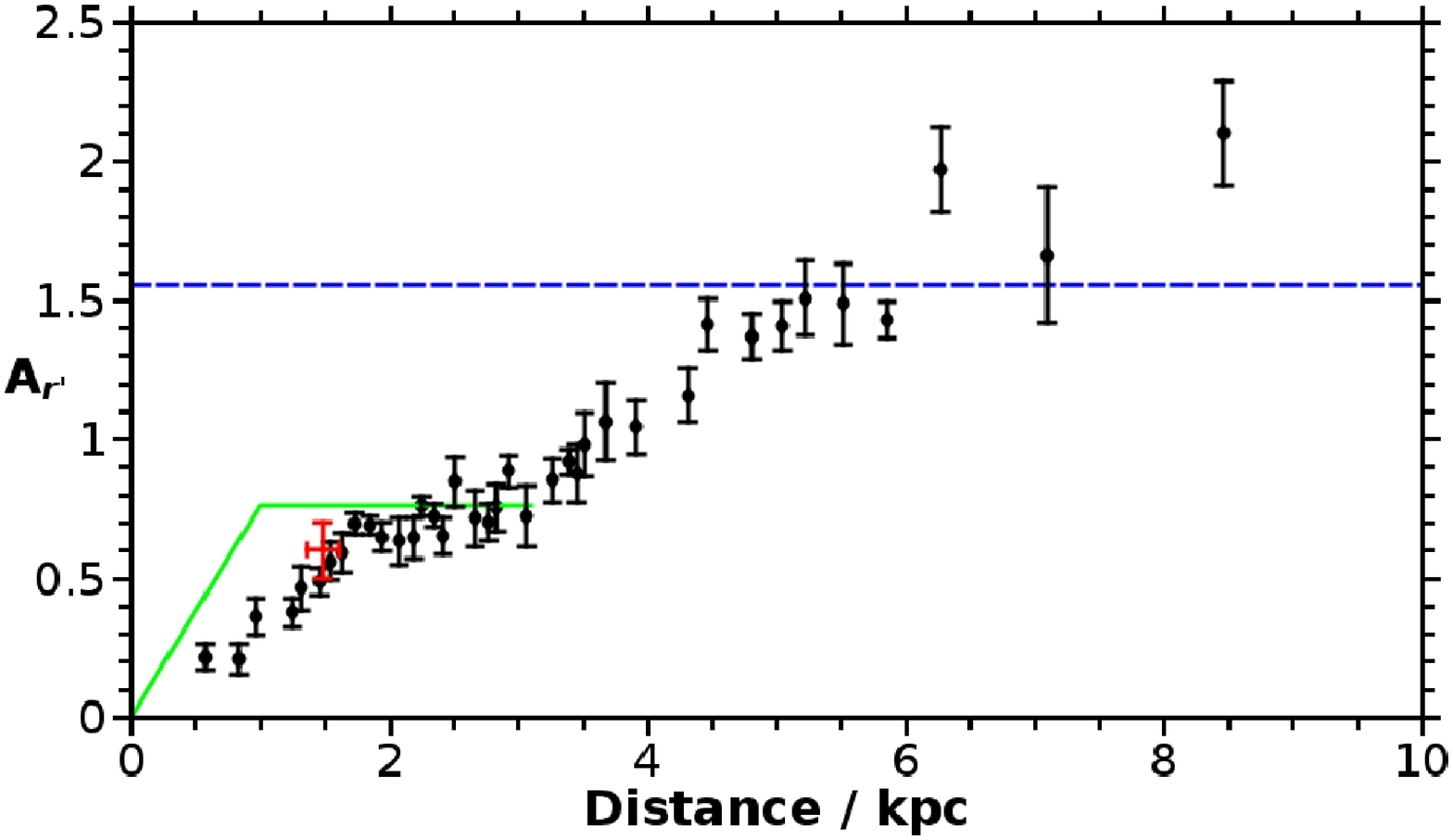}
\caption{The retrieved extinction-distance relationship for $(l,b)=(177.6, 3.1)$ (field 2849), for a $10 \arcmin \times 10 \arcmin$ box. The distance-extinction relationship derived in this work is in black, with the error bars indicated the errors on the derived mean value of extinction in each bin. The result of \protect\cite{Neckel.1980} is in green, the \protect\cite{Schlegel.1998} asymptotic value is shown with the blue dashed line and the distance and extinction of  NGC 2099 as found by \protect\cite{Hartman.2008} is shown in red. Estimates of extinction and distance for 752 objects have been binned up to produce the distance-extinction relationship.}
\label{2849_1}
\end{figure}

\begin{figure}
\centering
\includegraphics[width=80mm, height=60mm]{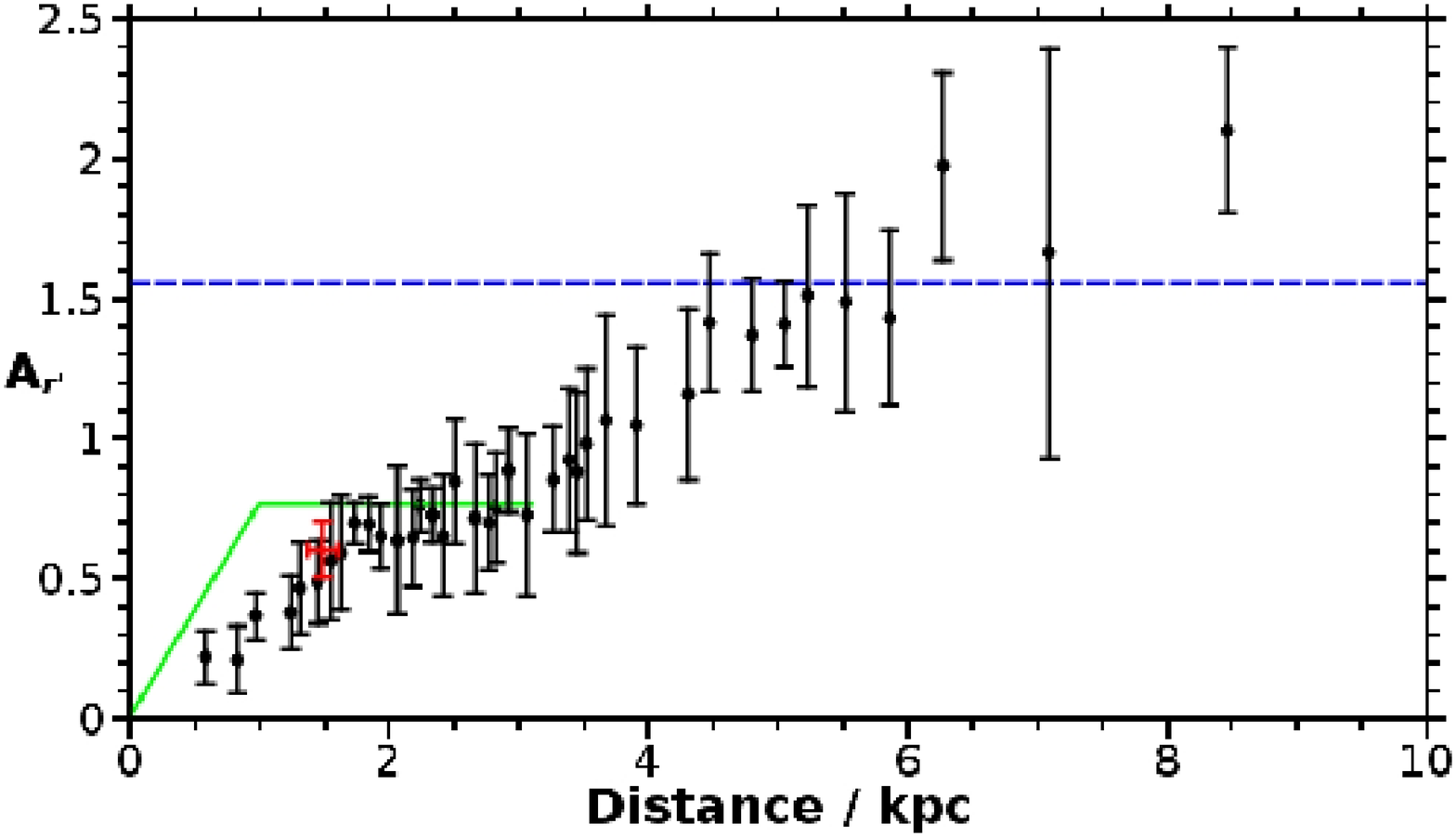}
\caption{As with Fig.~\ref{2849_1}, only here the error bars indicate the estimated intrinsic variation of extinction in the bin. \label{2849_2}}
\end{figure}

The Galactic cluster, NGC 2099 (M37) lies near the Anticentre at $(l,b)=(177.6, 3.1)$). \cite{Hartman.2008} measure this cluster to be at a distance of $1.49 \pm 0.12$~kpc with a reddening of $E(B-V)=0.23 \pm 0.04$. This sightline falls within IPHAS field 2849.  The distance-extinction relationship derived in this work and in the literature, are shown in Figures~\ref{2849_1} and \ref{2849_2}. The difference between the two plots is that in the first the errors on the estimated value of extinction for each bin are marked (these arise from systematic and random errors involved in obtaining the photometry), while in the second, the error bars reflect the intrinsic scatter that is a consequence of the existence of unresolved substructure in the interstellar medium. Both of these must be known to estimate the distance of an object based on its extinction and the distance-extinction relationship. Using the distance-extinction relationship shown above and the \cite{Hartman.2008} estimate of reddening, the distance to NGC 2099 is estimated to be $2.0 \pm 0.9$~kpc, where the large error is a result of the extinction-distance relationship flattening off between $\sim 2$~kpc and $\sim 3$~kpc. Based on the distance to W3OH of $1.95 \pm 0.04$~kpc \citep{Xu.2006}, this void in extinction would appear to lie directly behind the Perseus spiral arm.

\cite{Nilakshi.2002} measure a reddening of $E(B-V)=0.30 \pm 0.04$ and a distance of $1.36 \pm 0.1$~kpc to NGC 2099, this estimate is a significantly poorer fit to derived distance-extinction relation than the estimate of \cite{Hartman.2008}.  The cause of the difference between these two results appears to be the adopted metallicity: \cite{Hartman.2008} adopt a nearly solar value, while \cite{Nilakshi.2002} prefer a reduced metallicity of $Z = 0.008$. If all objects were assumed to have fainter absolute magnitudes, in line with the \cite{Nilakshi.2002} metallicity, the distance and extinction determined by \cite{Hartman.2008} remains a better fit to the produced distance-extinction relation.

\subsection{Cygnus OB2}\label{6010}

\begin{figure}
\centering
\includegraphics[width=80mm, height=60mm]{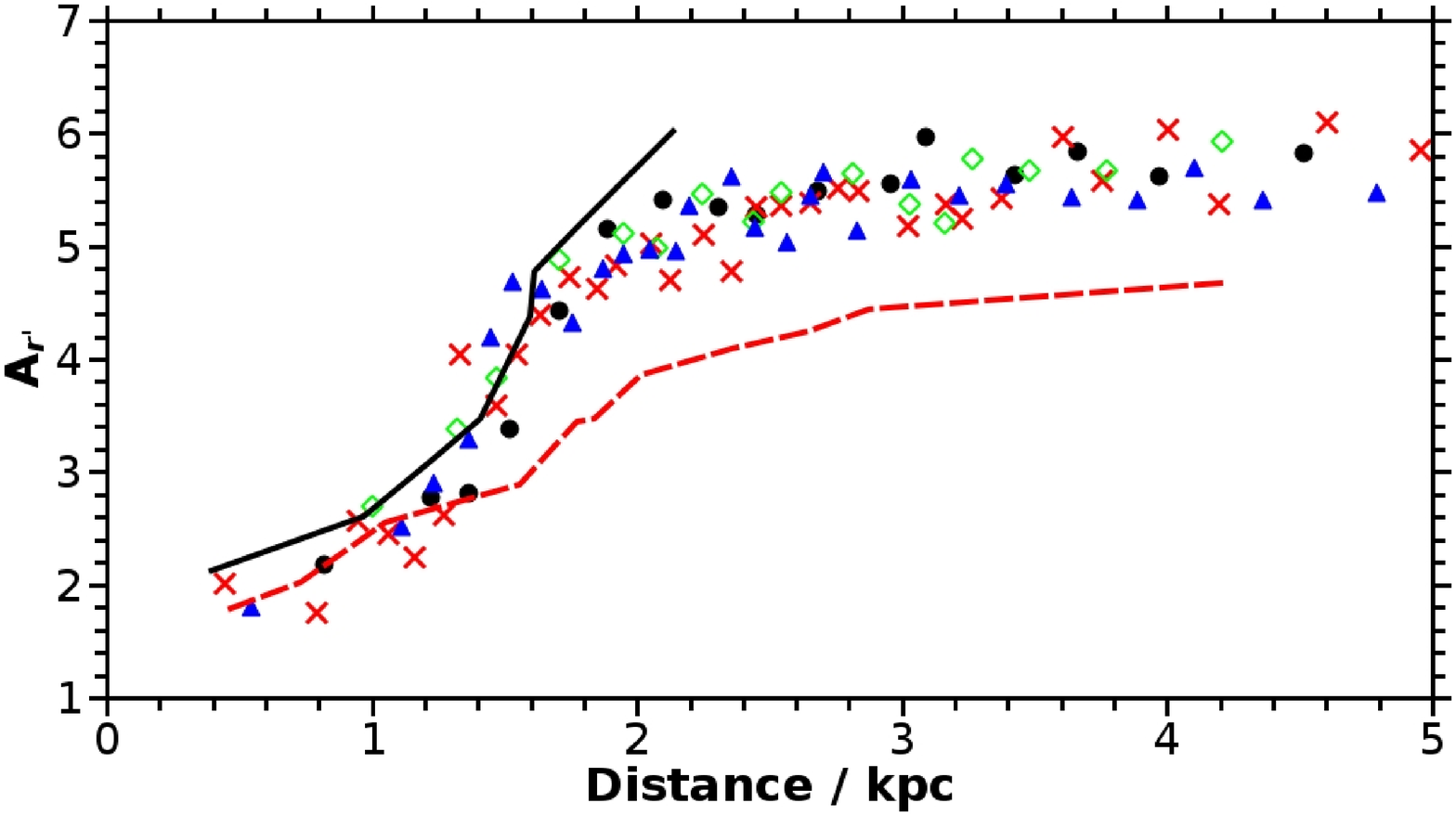}
\caption{Distance-extinction relationships for the IPHAS field 5985, in Cyg OB2, which has been divided into four quadrants. The relationship for the south-east quadrant is indicated by black dots, the south-west by red crosses, the north east by green open diamonds and the north west by blue triangles. Also shown are the mean distance extinction curves for fields 6010 (black solid line) and 6134 (red dashed line). \label{5985_1}}
\end{figure}

A potentially difficult environment for {\scshape mead} to operate in is one with substantial on-going star formation (Section~\ref{young_pops_test}). In such regions there will be relatively  large numbers of pre-main sequence objects which are not explicitly identified in the algorithm and therefore may be more subject to misclassification, and in addition any O and B stars present will be mistaken for A/F stars. One such region is the Cygnus OB2 association. Cyg~OB2 has been extensively studied in the past, and in particular \cite{Drew.2008} utilised IPHAS photometry to study the positions of early A-stars around the association, concluding that it lies at a distance between $1.45$~kpc and $1.75$~kpc, with the range in distances a result of uncertainty in the age of the association.

IPHAS fields 5985 and 6010 look towards the approximate centre of the Cygnus OB2 association, having field centres at $(l,b)=(80.3, 1.2)$ and $(l,b)=(80.2, 0.7)$ respectively. Figure~\ref{5985_1} show the derived distance-extinction relationships for these fields: in both cases a substantial increase in extinction is apparent at $\sim1.5$~kpc. This rise in extinction would therefore appear to be associated with Cygnus OB2 association. In field 5985 the extinction flattens out and approaches its asymptotic value. This flattening is not seen in the results for field 6010 -- most likely due to the extinction continuing to rise significantly beyond Cyg OB2, placing most objects outside the magnitude limits of the observations. \cite{Schlegel.1998} find significantly higher values of asymptotic extinction in field 6010 than in 5985, consistent with this \citep[see also][]{Drew.2008}.

The range over which the extinction sharply rises in field 5985 is $\sim 400$~pc. It is not likely that the extinction associated with the Cygnus OB2 association itself is this deep, but rather that the cumulative errors are inducing a smoothing of the distance-extinction relationship \citep[a point commented on by][]{Juric.2008short}. Using the Monte-Carlo Method for simulating photometry, it is possible to produce a result similar to the ones displayed in both fields with the extra extinction confined to a depth of $\la 100$~pc at a distance of $(1.5 \pm 0.2)$~kpc in the model.

\begin{figure}
\centering
\includegraphics[width=80mm, height=60mm]{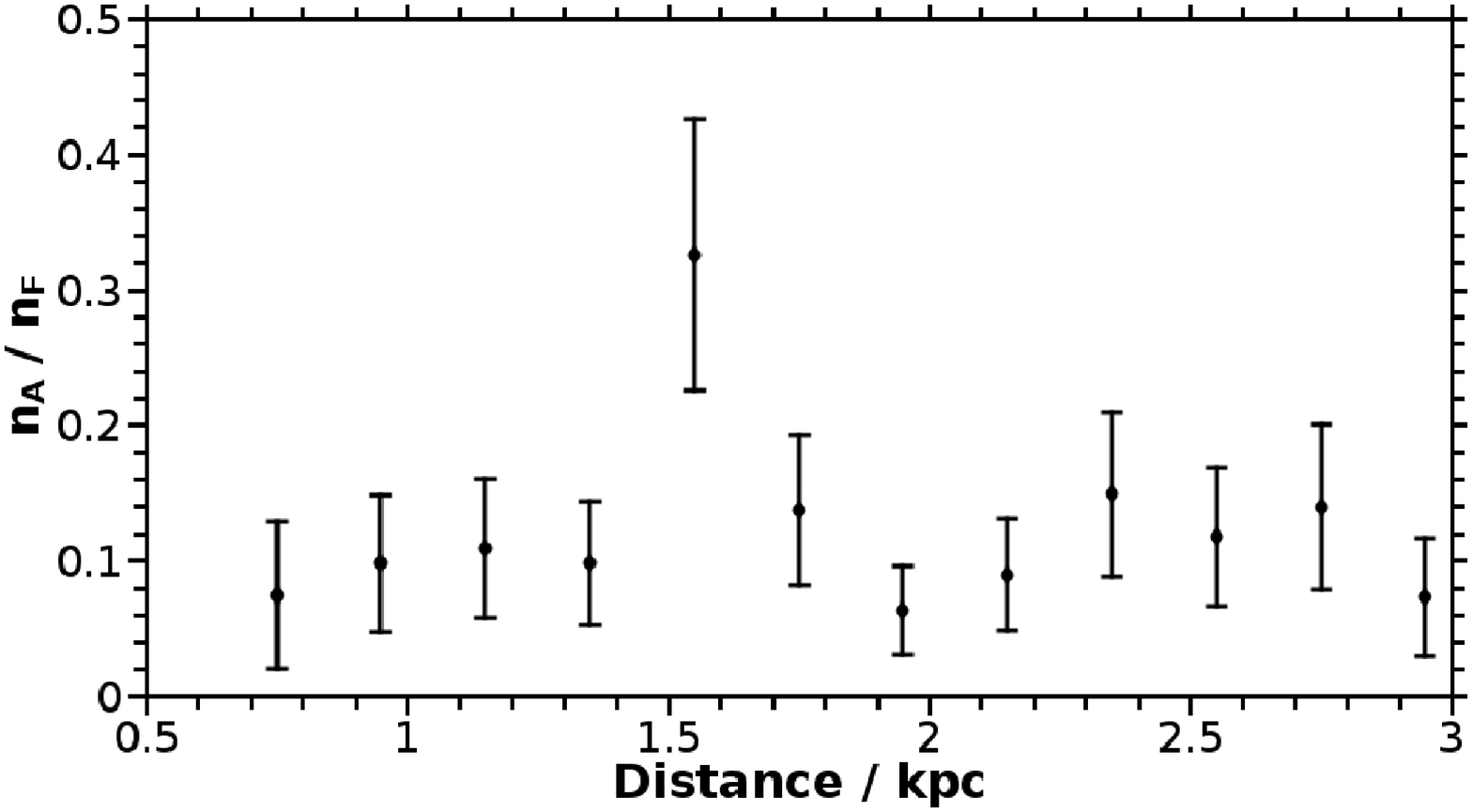}
\caption{The ratio of the numbers of near Main Sequence early-A stars to F dwarfs as a function of distance for the field 5985.\label{5985_2}}
\end{figure}

A second method for determining the distance to the Cygnus OB2 association is to determine the ratio of the number of younger stars to a sample with longer lifetimes along the line of sight. Early A (A0--5) near Main Sequence stars were used as the younger population due to their relative ease of selection and the low probability of contamination by non-MS stars. Their number was compared with that of late F-type dwarfs (F5--9), a population with a main sequence lifetimes roughly $10$ times longer. Ideally a sample of redder dwarfs or giants would be preferred in this role, but later-type dwarfs are too faint to be observed over a large enough distance range, due to the rapidly-rising extinction, while giants are not sufficiently numerous in this region. Fig.~\ref{5985_2} demonstrates the dependence of this ratio on distance within field 5985, showing a significant peak at $1.55$~kpc. This distance is consistent with both that determined in \cite{Drew.2008} and the rise in extinction shown in Fig.~\ref{5985_1}. The peak shown in Fig.~\ref{5985_2} is not as broad as the region of high differential extinction in Fig.~\ref{5985_1}, this is a consequence of the near Main Sequence early-A stars being relatively bright and so exhibiting smaller photometric and therefore distance, errors.  

As previously mentioned, a significant concern when analysing fields in Cyg~OB2 is contamination by PMS objects. PMS objects will exhibit brighter absolute magnitudes than MS objects of the same intrinsic $(r'-i')$ colour and thus their distance would be underestimated if they were classified as dwarfs. Therefore it might be expected that distance bins closer than Cyg~OB2 might incorrectly contain objects with higher extinctions, thus causing the estimate of mean extinction in these bins to be too high. However, many PMS objects will actually be classified in luminosity classes IV or III, especially if they have blue intrinsic colours where the luminosity classes are tightly spaced (Fig.~\ref{HRD}). The ability to classify these objects in classes IV or III will result in a more accurate estimation of distance. But it must be acknowledged that there is room for some falsification through application of absolute magnitude scales derived for evolved stars.  A second, potentially more serious, problem is a result of the behaviour of PMS objects on the colour-colour diagram, particularly in relation to the $\Halpha$ band. Although substantial $\Halpha$ line emission will move an object outside the main stellar locus, less pronounced emission could result in an object still lying within the main stellar locus and {\scshape mead} estimating an intrinsic colour redder than reality and too low a value of extinction. 

Field 6134 ($(l,b)=(81.46, -0.72)$) is outside Cyg~OB2, but due to its relative proximity should exhibit similar properties to fields 5985 and 6010 at distances less than that to Cyg~OB2. Fig~\ref{5985_1} demonstrates that the local ($<1$~kpc) behaviour of extinction in fields 5985, 6010 and 6134 is much the same. This is as it should be, since at these near distances the field centres are less than 50~pc apart and likely to be sampling much the same stellar environment. Further examination of the results for field 5985 reveals a $30\%$ increase in the proportion of objects in luminosity classes IV and III, relative to field 6134, suggesting that {\scshape mead} is indeed capable of dealing with the brighter absolute magnitudes of PMS objects. 

\section{Discussion}

We have presented an algorithm, {\scshape mead}, and shown how it can be applied to the $r'$, $i'$ and narrow-band $H\alpha$ photometry forming the IPHAS database to map out extinction in the Galactic disc at high angular and depth resolution.  The algorithm is fully empirical in that it depends only on absolute magnitude calibrations (especially the Houk et al 1997 HIPPARCOS main sequence) and not on any Galactic model. Because of this it is better able to deal with perturbations away from a smooth structure. This is crucial as it is these perturbations which are of real interest, in view of their possible origin due to either accreted satellite galaxies or to the Milky Way's still-debated spiral structure.

\cite{Juric.2008short} are able to successfully correct for extinction using the 2D maps due to \cite{Schlegel.1998}, since the majority of their observations sample high Galactic latitudes. For studies of the Galactic disc at low latitudes, a more sophisticated treatment of extinction is required. The algorithm {\scshape mead} presented here is a means to achieving this.  While the \cite{Juric.2008short} SDSS sample is effectively volume limited for most spectral types, leading to main sequence stars dominating the sample, the IPHAS database is more clearly magnitude limited, requiring the disentangling of luminosity class.  This capability is a critical new feature of {\scshape mead}. 

\begin{figure}
\centering
\includegraphics[width=80mm, height=60mm]{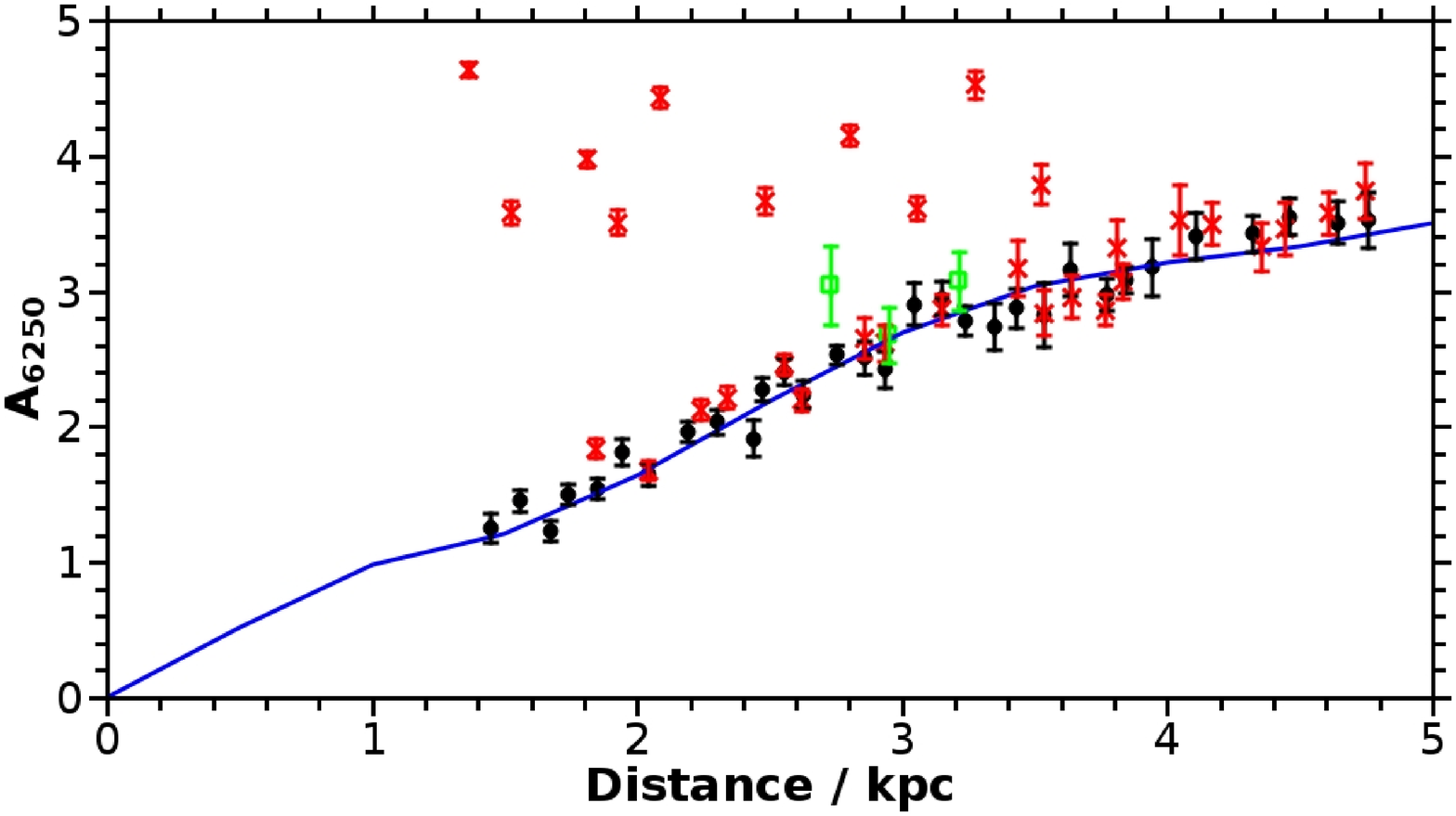}
\caption{The retrieved extinction-distance relationship for a $10 \arcmin \times 10 \arcmin$ box in the direction $(l,b)=(180,0)$  using the final version of {\scshape mead} (black dots), near MS early-A stars only (green open squares) and by assuming that all objects are dwarfs (red crosses). \label{detail_compare}}
\end{figure}

The new power of {\scshape mead} can be illustrated by comparing the distance-extinction relationships obtained (i) using A stars only \citep[as in][]{Drew.2008}, (ii) analysing intrinsic colour, whilst classifying all stars as dwarfs, (iii) using {\scshape mead} in full (Fig.~\ref{detail_compare}).  \cite{Drew.2008} demonstrated that extinctions and distances of near Main Sequence early-A stars were particularly straightforward to derive from their IPHAS colours and apparent magnitude. This approach, on its own, could in principle be used to map extinction, but near Main Sequence early A-stars represent only a small proportion of objects in the IPHAS catalogue. By accessing a broader range of intrinsic stellar colours  {\scshape mead} is able to map extinctions on finer spatial scales, with greater accuracy and over a wider distance range.  This figure also demonstrates why it is so important to analyse luminosity class as well as intrinsic colour.

As well as estimating the distance-extinction relation, {\scshape mead} also returns distance, intrinsic colour and extinction estimates for a large proportion of objects in the IPHAS database. Interpreting the resultant catalogue, will be non-trivial, but will enable the structure of the Galactic plane to be studied in a manner similar to the analysis of the Milky Way at higher latitudes by \cite{Juric.2008short}. For example, scale heights and lengths of both the dust and stars within the disc will be derivable.  

A side benefit of {\scshape MEAD} is that it generates lists of objects that do not conform to the most common spectral classifications.  These 'rejected' objects may still be of interest for focused studies of rare phases of stellar evolution.  Whilst, most of the objects in this category will be late K or M-type stars, or noise dominated, others would be extreme red giants \citep{Wright.2008}, sub-luminous objects (Section~\ref{contaminents_test}) or emission line stars. \cite{Witham.2008} have already presented a preliminary catalogue of emission line stars extracted from above the main stellar locus.

In the future, {\scshape mead} could also be applied to the results of the upcoming VST/OMEGACAM Photometric Hydrogen Alpha Survey of the Southern Galactic Plane (VPHAS+). Combining the results from VPHAS+ and IPHAS would lead to an extinction map of the entire Galactic plane in unprecedented detail. VPHAS+ and UVEX extend the wavelength coverage of IPHAS to the $u'$ and $g'$ bands over the full plane.  These additional bands open up the investigation of variations in metallicity \citep[cf.][]{Ivezic.2008short} and in the extinction law.  Here, in limiting ourselves to red wavelengths, we have been able to work with a standard $R = 3.1$ law as reasonably representative of most sightlines.  Combination with near-infrared wavelengths is also becoming an option as the UKIDSS GPS progresses and VISTA (Visible and Infrared Survey Telescope for Astronomy) begins data collection.  

As a next step, the impending completion of IPHAS observing creates the opportunity to apply {\scshape mead} to a full 3D mapping of the northern Galactic Plane (Sale et al. in preparation). 

\section*{Acknowledgments}

This paper makes use of data obtained as part of the INT Photometric H$\alpha$ Survey of the northern Galactic Plane (IPHAS) carried out at the Isaac Newton Telescope (INT). The INT is operated on the island of La Palma by the Isaac Newton Group in the Spanish Observatorio del Roque de los Muchachos of the Instituto de Astrofisica de Canarias. All IPHAS data are processed by the Cambridge Astronomical Survey Unit, at the Institute of Astronomy in Cambridge. SES is in receipt of a studentship funded by the Science \& Technology Facilities Council of the United Kingdom. DS acknowledges support through an STFC Advanced Fellowship.

This research has made use of the VizieR catalogue access tool and the SIMBAD database, both operated at CDS, Strasbourg, France. This research has made use of NASA's Astrophysics Data System Bibliographic Services

\bibliography{main}

\end{document}